\documentclass[12pt]{iopart}

\usepackage{amsmath}
\usepackage{bm}
\usepackage{xcolor}
\usepackage{hyperref}
\usepackage[english]{babel}
\usepackage{subfig}
\usepackage{caption}
\usepackage{tabularx}
\usepackage{booktabs}
\usepackage{iopams}
\usepackage{graphicx}
\begin{document}

\title[Anomalous flux periodicity in proximitised quantum spin Hall constrictions]{Anomalous flux periodicity in proximitised quantum spin Hall constrictions}

\author{Lucia Vigliotti$^1$, Alessio Calzona$^3$, Bj\"orn Trauzettel$^{2,3}$, Maura Sassetti$^{1,4}$, Niccolò Traverso Ziani$^{1,4}$}

\address{$^1$Dipartimento di Fisica, Università degli Studi di Genova, Via Dodecaneso 33, 16136, Genova, Italy}
\address{$^2$Institute for Theoretical Physics and Astrophysics,
University of W\"urzburg, D-97074 W\"urzburg, Germany
}
\address{$^3$W\"urzburg-Dresden Cluster of Excellence ct.qmat, Germany}
\address{$^4$CNR SPIN, Via Dodecaneso 33, 16136, Genova, Italy}
\ead{traversoziani@fisica.unige.it}
\vspace{10pt}
\begin{indented}
\item[]
\end{indented}

\begin{abstract}
We theoretically analyse a long constriction between the helical edge states of a two-dimensional topological insulator. The constriction is laterally tunnel-coupled to two superconductors and a magnetic field is applied perpendicularly to the plane of the two-dimensional topological insulator. The Josephson current is calculated analytically up to second order in the tunnel coupling both in the absence and in the presence of a bias (DC and AC Josephson currents). We show that in both cases the current acquires an anomalous $4\pi$-periodicity with respect to the magnetic flux that is absent if the two edges are not tunnel-coupled to each other. The result, that provides at the same time a characterisation of the device and a possible experimental signature of the coupling between the edges, is stable against temperature. The processes responsible for the anomalous $4\pi$-periodicity are the ones where, within the constriction, one of the two electrons forming a Cooper pair tunnels between the two edges.
\end{abstract}

%
%
%
%
%

\section{Introduction}
The quest for ballistic electronic channels in the absence of magnetic fields pushed intense scientific efforts into the development of topologically protected edge states beyond the quantum Hall effect \cite{ti1,ti2}. Based on the theoretical proposals by Kane and Mele \cite{km1,km2} and by Bernevig, Hughes, and Zhang \cite{bhz}, in 2007 such edge states were detected in HgTe-CdTe heterostructures \cite{exp1,exp2}. The state of electronic matter that was created is called quantum spin Hall effect (QSHE). In the following years many other platforms added up to HgTe-CdTe heterostructures, ranging from InAs-GaSb heterostructures \cite{inas1,inas2}, to bismuthene \cite{bis}, 1T'-WTe2 \cite{mono}, and jacutingaite \cite{brasil}.

The QSHE is characterised by the fact that at low energy the electrons circulate around the edges, as in the quantum Hall effect. However, interestingly, they have well defined helicity, meaning that electrons with opposite spin projection move in opposite directions \cite{km1,km2,helical1,helical2}. This fact opens unprecedented possibilities in spintronics \cite{spin1,spin2,spin3,spin4}. Additional technological applications of the QSHE are in topologically protected quantum computation \cite{tpc1,tpc2,tpc3,tpc4}, through the engineering of Majorana fermions \cite{majo1,majo2,majo3,majo4,majo5,majo6,majo7,majo8,majo9} and parafermions \cite{para1,para2,para3,para4}, and in superconducting spintronics \cite{sspin1,sspin2,sspin3,sspin5,sspin4}. At a more fundamental level, electron quantum optics experiments \cite{eqo1,eqo2,eqo3} and fractional soliton physics \cite{frac1,frac2,frac3,frac4,frac5} could be inspected.

To fully exploit the potential of the QSHE, nanostructuring must be performed in order to manipulate the edge states. In this direction, a substantial effort has been devoted to the proximisation with superconductors \cite{sc1,sc2,sc3,sc4} and to the creation of constrictions between the edges \cite{nat,sqpc1,sqpc2}. Indeed, the combination of the two could allow - in complex setups - for the creation of Majorana fermions, parafermions, Floquet bound states, and equal spin pairing \cite{qpc1,qpc2,qpc3,qpc4,qpc5,qpc6}. Thermal and thermoelectric properties have also been addressed \cite{ttm1,ttm2}. Most remarkably, the engineering of long constrictions (on the scale of the inverse Fermi momentum) makes ferromagnetic barriers, that were never implanted on QSHE systems, unnecessary. However, surprisingly, the main building block of the aforementioned proposals, that is the long topological constriction with superconducting contacts, is still largely unexplored from the theoretical point of view and has never been experimentally realised. Moreover, the only result reported in literature dealing with a long topological constriction, in the absence of superconductivity, is about an interaction induced reduction of the conductance \cite{nat}, that could disappear in the presence of superconductors, due to screening. It is hence even difficult to pinpoint a clear signature of the formation of the constriction in the presence of superconductors.

In this work, we calculate the current-voltage relation for the long constriction with superconducting contacts and, at the same time, we provide an accessible scheme for demonstrating the formation of topological constrictions in the presence of superconductors. More specifically, we characterise theoretically the Josephson current in presence of a magnetic field perpendicular to the plane defined by the edges. What we find is that, due to the constriction, the current acquires a $4 \pi$-periodicity in the flux related to the magnetic field. Such a periodicity is visible both in the zero bias and in the finite bias regime. We also show that the signature is robust - or even increased - at nonzero temperature. It is here worth noticing that the anomalous periodicity we find is not strictly related to crossings in the Andreev bound states and Majorana physics \cite{tpc1}. Anomalous periodicities have been reported in \cite{buzdin}, but based on different physics in a Josephson junction with a nanowire as non-superconducting element and not a quantum spin Hall constriction.

The rest of the article is structured as follows: in Sec. \ref{sec2} we present our model; in Sec. \ref{sec3} we outline the formalism we use for the calculation of the transport properties. Sec. \ref{sec4} is devoted to the presentation of the results, and finally in Sec. \ref{sec5} we draw our conclusions and discuss the experimental parameters. Several technical details are presented in the Appendices.

\section{Model}
\label{sec2}
\begin{figure}[h!]
	\centering
	\includegraphics[width=10cm]{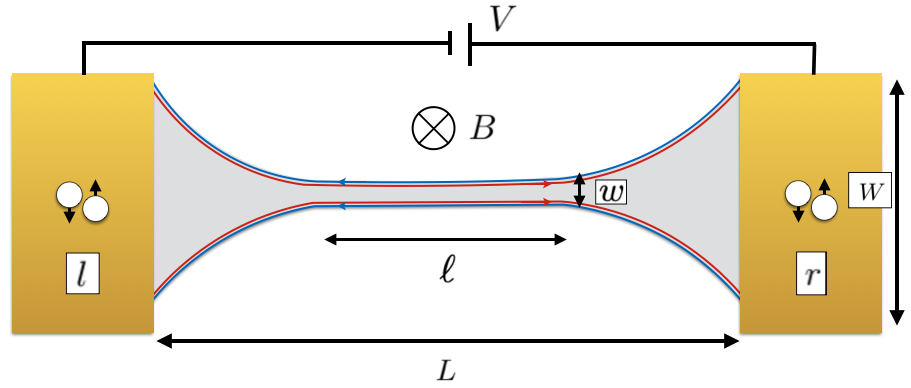}
	\caption{\small{Schematic of the setup: a sample of 2DTI of length $L$ and width $W$ is laterally tunnel-coupled to two superconductors (right, $r$ and left, $l$); a magnetic field $B$ applied perpendicularly to the plane of the 2DTI and a bias $V$ are also included. Halfway in the 2DTI region, there is a constriction of length $\ell$ and width $w$ between the helical edges.}}
	\label{fig:setup}
\end{figure}

The figure below (Fig. \ref{fig:setup}) shows the setup we are going to inspect, consisting of a Josephson junction made of a two-dimensional topological insulator (2DTI) sample of length $L$ and width $W$ tunnel-coupled to two superconductors (SCs), right and left. The superconducting part is a proximitised region of the topological structure. Halfway in the 2DTI region, a constriction between the helical edge states, with length $\ell$ ($\ell<L$, $\ell>{k_F}^{-1}$, with $k_F$ the Fermi momentum) and width $w\ll W$, is present. A magnetic field $B$ applied perpendicularly to the plane of the 2DTI and a bias $V$ are also included.
In the following we are going to analyse one by one all the mentioned elements.

\subsection{The constriction}
\begin{figure}[h!]
	\centering
	\includegraphics[width=10cm]{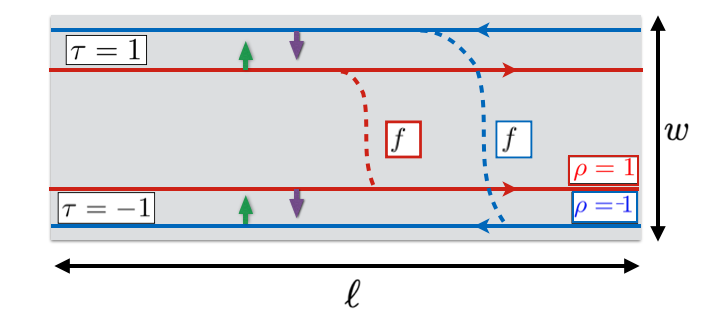}
	\caption{\small{Schematic of the constriction of length $\ell$ and width $w$. The direction-conserving couplings between the upper and lower edge of the 2DTI, with amplitude $f$, are also shown.}}
	\label{fig:tunnelling}
\end{figure}

The basic ingredient is represented by the constriction between the helical edges of the 2DTI. They extend for $-L/2<x<L/2$ and are $W$ apart from each other at the interfaces with the SCs, while in the narrow constriction of length $\ell< L$, $\ell>k_F^{-1}$, their separation reduces to $w\ll W$ (see Fig. \ref{fig:tunnelling}).
As $L\gg k_F^{-1}$, the system under inspection acquires translational invariance and the momentum $k$ becomes a good quantum number. The associated Hamiltonian $\hat{H}_E^0$ is given by 
\begin{equation}\label{eqn:H_E^0}
\hat{H}_E^0=\sum_k \bm{\hat{c}}^{\dagger}_k\mathcal{H}_E^0\bm{\hat{c}}_k.
\end{equation}
Here, we have
\begin{equation}
\bm{\hat{c}}_k=\left(\hat{c}_{k,11},\hat{c}_{k,-11},\hat{c}_{k,-1-1},\hat{c}_{k,1-1}\right)^T,
\end{equation}
with $\hat{c}_{k,\rho\tau}$ the Fermi operator that annihilates an electron with momentum $k$ propagating in the $\rho$-direction channel of the $\tau$ edge. We set $\rho=1(-1)$ for the right (left) direction of motion and $\tau=1(-1)$ for the upper (lower) edge. Due to the spin-momentum locking, these two indices completely define the edge states, since the spin polarisation is determined by their helical nature, as shown in Fig. \ref{fig:tunnelling}. Moreover, we define
\begin{equation}
\mathcal{H}_E^0=\mathcal{H}_{kin.}+\mathcal{H}_{f.s.},
\end{equation}
with
\begin{equation}
\mathcal{H}_{kin.} = \hbar v_Fk\tau_3\otimes\rho_3-\mu\tau_0\otimes\rho_0 \label{eqn:kin}
\end{equation}
representing the kinetic energy with $v_F$ the Fermi velocity and $\mu$ the chemical potential. In the equation above, $\rho_i$ and $\tau_i$, $i=0,1,2,3$ represent the identity and the three Pauli matrices acting on right/left mover and upper/lower edge space, respectively. Furthermore,
\begin{equation}
    \mathcal{H}_{f.s.}=f\tau_1\otimes\rho_1\label{eqn:forward}
\end{equation}
describes a forward scattering tunnelling term across the edges parametrised by $f$. More specifically, it represents the processes where one electron changes the edge while preserving the direction of motion, and hence flips its spin. Note that the Hamiltonian is time-reversal invariant. A remark is here in order: according to the spatial separation of the helical edges, we assume the tunnelling to take place only in the constriction, where $w\ll W$. However, in the long constriction case we inspect, we only consider tunnelling events that conserve momentum $k$, meaning that we neglect finite size effects related to $\ell$.

A backward scattering term of the type $\mathcal{H}_{b.s.}= b\tau_1\otimes\rho_0$ is not considered here, as well as any gapping of the edges due to intra-edge mechanisms; indeed, backscattering (and small gaps opened by intra-edge mechanisms) do not significantly affect the system away from the Dirac point. Since the latter is located, in the thick heterostructure case, deep in the valence band \cite{nat}, we can ignore these effects. They could be included perturbatively but we do not expect them to affect our results.\\
Supported by the coherent transport over long distances in state of the art samples, we assume the absence of impurities \cite{spin4}. For a schematic of the system and of the couplings we consider, see Fig. \ref{fig:tunnelling}.

Diagonalising Eq. (\ref{eqn:H_E^0}), we obtain
\begin{equation}
\hat{H}_E^0=\sum_{i=1}^4\sum_k E_{A_i}(k)\hat{A}_{k,i}^{\dagger}\hat{A}_{k,i},\label{eqn:H_E^0A}
\end{equation}
with
\begin{eqnarray}
	\hat{A}_{k,1}&=&\frac{1}{\sqrt{2}}\left(-\hat{c}_{k,-11}+\hat{c}_{k,-1-1}\right)\notag\\
	\hat{A}_{k,2}&=&\frac{1}{\sqrt{2}}\left(\hat{c}_{k,-11}+\hat{c}_{k,-1-1}\right)\notag\\
	\hat{A}_{k,3}&=&\frac{1}{\sqrt{2}}\left(-\hat{c}_{k,11}+\hat{c}_{k,1-1}\right)\notag\\
	\hat{A}_{k,4}&=&\frac{1}{\sqrt{2}}\left(\hat{c}_{k,11}+\hat{c}_{k,1-1}\right),
\label{eqn:En.A}
\end{eqnarray}
and
\begin{eqnarray}
E_{A_1}(k)&=& -f-\hbar v_F k-\mu\notag\\
E_{A_2}(k)&=& f-\hbar v_F k-\mu\notag\\
E_{A_3}(k)&=& -f+\hbar v_F k-\mu\notag\\
E_{A_4}(k)&=& f+\hbar v_F k-\mu.
\label{eqn:En.A2}
\end{eqnarray}
These four eigenstates have well-defined chirality (left for $\hat{A}_{k,1}$, $\hat{A}_{k,2}$ and right for $\hat{A}_{k,3}$, $\hat{A}_{k,4}$). We highlight the $1/\sqrt{2}$ weight in each of the combination $\hat{A}_{k,i}$, which means that the new eigenstates are an equal superposition of an upper-edge state and a lower-edge state. The role of $f$ is hence to split, in energy, the dispersion. Notice the analogy with the Rashba coupling in quantum wires \cite{rw1,streda,rw2,rw3,rashba}.

In the following, when discussing the tunnelling processes, we will also introduce a magnetic flux $\phi$ through the plane of the 2DTI.

\subsection{Superconducting leads and effective Hamiltonian}
\label{subsec:2.2}
The edges just described are proximitised, for $x<-L/2$ and $x>L/2$, with standard BCS superconductors. The left SC, extending for $x<-L/2$, is indexed by $j=-1$ while the right one, extending for $x>L/2$, by $j=1$. The superconducting pairing is denoted by $\Delta$ and the bare superconducting pairing phases - that we keep distinguished in principle - by $\varphi_j^0$. Moreover, we assume the chemical  potential to be the same in the two SCs. Each SC is described by a Hamiltonian $\hat{H}^j_S$ and the coupling between the SCs and the constriction is modeled by means of a time-reversal invariant Hamiltonian \cite{loss} $\hat{H}_T=\sum_{j}\hat{H}_T^j$. The explicit form of $\hat{H}^j_S$ and $\hat{H}^j_T$ can be found in Appendix A.

\begin{figure}[h!]
	\centering
	\includegraphics[width=9cm]{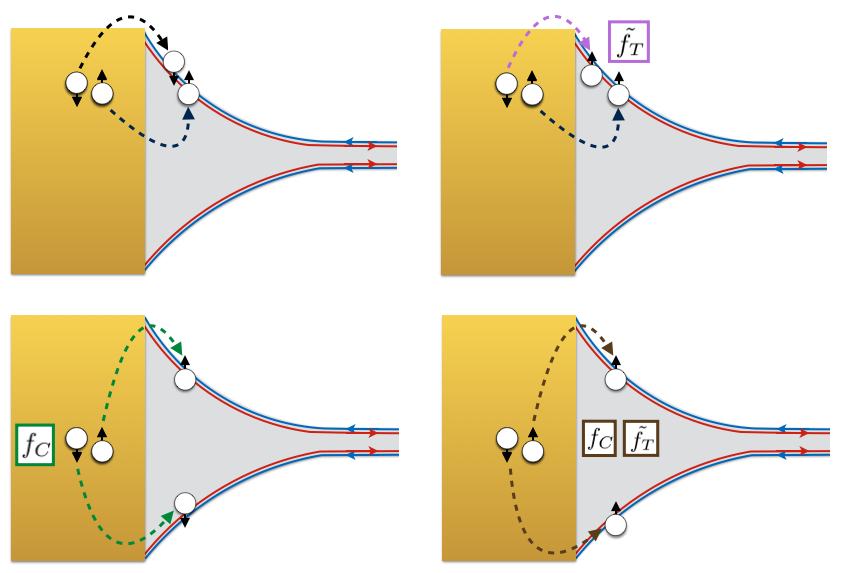}
	\caption{\small{Possible processes of injection of two electrons: in a spin-singlet or spin-triplet state (left or right panels) and either into the same edges or into different ones (upper or lower panels).}}
	\label{fig:tunnelproc}
\end{figure}

To obtain an effective Hamiltonian of the proximitised system, we integrate out the SCs. This calculation is performed in direct space and in the absence of the magnetic field. Eventually, we reinsert the magnetic field keeping in mind the gauge invariance of the phase \cite{loss}.\\
The extra term appearing in the edge Hamiltonian can be approximately written as \cite{loss}
\begin{equation}
\delta \hat{H}_E\approx \sum_{\zeta_1,\zeta_2,j}\left[\Gamma_{\zeta_1,\zeta_2,j}\hat{\psi}_{\zeta_1}(x_j^-)\hat{\psi}_{\zeta_2}(x_j^+)+h.c.\right],\label{eqn:dH}
\end{equation}
with $\hat{\psi}_{\rho\tau}(x)=1/\sqrt{L}\sum_k \hat{c}_{k,\rho\tau} e^{ikx}$. In Eq. (\ref{eqn:dH}), $\zeta_1,\zeta_2$ are collective indices standing for $\rho_1\tau_1,\rho_2\tau_2$, and $x_j^{\pm}=jL/2\pm\delta_{\zeta_1,\zeta_2}\xi/2$, where $\xi=\hbar v_F/\Delta$ is the coherence length in the edges, which represents the short distance cutoff of our system, $\xi\ll L$. The approximation is done for the regime $E\ll\Delta$. In this regime, transport between the SCs and the edges is essentially carried out by Cooper pairs (CPs) and there is no contribution of single quasi-particles.\\
Eq. (\ref{eqn:dH}) accommodates every possible process of injection of the two electrons (see Fig. \ref{fig:tunnelproc}): either in a spin-singlet or spin-triplet state (the triplet processes being proportional to a factor $\tilde{f}_T=f_T/\sqrt{1+f_T^2}$\footnote{$f_T$ is the ratio of spin-flipping tunnelling processes over spin-conserving ones. It is reasonable to include such a parameter in the model since it allows to take into account the Rashba coupling in the material, which makes spin flips possible. Typically, $f_T\ll1$.}); either into the same edges (``direct Andreev reflection", DAR) or into different ones (``crossed Andreev reflection", CAR). CAR is possible as we assume $\xi_S>W$, where $\xi_S=\hbar v_{F,S}/\Delta$ (with $v_{F,S}$ the Fermi velocity in the SCs) is the coherence length of the SCs, and its suppression with respect to DAR will be denoted by $f_C$\footnote{Here, $f_C\sim \mathsf{f}(k_{F,S}W)e^{-W/\xi_S}$, with $\mathsf{f}$ an oscillatory and decaying function depending on the spatial dimension of the SCs and $k_{F,S}$ the Fermi momentum in the SCs. This quantity naturally emerges while integrating out the SCs, see \cite{loss} for the details.}. Note that the splitting $x_j^{\pm}$ makes tunnelling of spin-triplet CPs into or out of the same edge possible.

The summation in Eq. (\ref{eqn:dH}) has been antisymmetrised for each $\zeta_1\neq\zeta_2$. Following this requirement, the summed terms are reduced to ten:
\begin{align*}
&\Gamma_{11,11,j}&\qquad&\Gamma_{11,-11,j}&\qquad&\Gamma_{11,-1-1,j}&\qquad&\Gamma_{11,1-1,j}&\qquad&\Gamma_{-11,-11,j}\\
&\Gamma_{-11,-1-1,j}&\qquad&\Gamma_{-11,1-1,j}&\qquad&\Gamma_{-1-1,-1-1,j}&\qquad&\Gamma_{-1-1,1-1,j}&\qquad&\Gamma_{1-1,1-1,j}.
\end{align*}
In momentum space, Eq. (\ref{eqn:dH}) becomes 
\begin{equation}
\delta \hat{H}_E= \sum_{\zeta_1,\zeta_2,j}\left[\Gamma_{\zeta_1,\zeta_2,j}\frac{1}{L}\sum_{k_1}\sum_{k_2}\hat{c}_{k_1,\zeta_1}e^{ik_1x_j^-}\hat{c}_{k_2,\zeta_2}e^{ik_2 x_j^+}+h.c.\right]=\delta \hat{H}^r_E+\delta \hat{H}^l_E.\label{eqn:dHk}
\end{equation}
Each of the $\Gamma_{\zeta_1,\zeta_2,j}$ coefficients contains all the details specifying the tunnelling process, i.e. species of the electrons forming the CP, spin-flipping, direct/crossed Andreev reflection. They are all proportional to the tunnelling rate $\Gamma=\pi \mathfrak{T}^2 N_S$, with $N_S$ the normal density of states per spin at the Fermi-level in the superconductors and $\mathfrak{T}$ the tunnelling coefficient related to the opacity of the barrier. We report the full expression for the $\Gamma_{\zeta_1,\zeta_2,j}$ coefficients, but not the whole calculation to obtain them, since they are a result already known in literature \cite{loss}\footnote{The minus sign in Eq. (\ref{eqn:gammaloss}), not present in \cite{loss}, is due to a different choice of the order for our edges basis, naturally leading to the definition of $\Gamma_{-1-1,1-1,j}\equiv\Gamma_{-1-1,1-1,j}-\Gamma_{1-1,-1-1,j}$ instead of $\Gamma_{1-1,-1-1,j}\equiv\Gamma_{1-1,-1-1,j}-\Gamma_{-1-1,1-1,j}$ as in the original paper.}:
\begin{equation}
\Gamma_{\zeta_1,\zeta_2,j}=(-1)^{\delta_{\zeta_1,-1-1}\delta_{\zeta_2,1-1}}\Gamma\left(\tilde{f}_T\right)^{\delta_{\rho_1*\tau_1,\rho_2*\tau_2}}(f_C)^{\delta_{\tau_1,-\tau_2}}e^{i[\frac{j}{2}k_FL(\rho_1+\rho_2)-\varphi^0_j]}.\label{eqn:gammaloss}
\end{equation}
As expected, the factor $f_C$ is present only if $\tau_1\neq\tau_2$ (CAR) and $\tilde{f}_T$ is present only if the two spins of the electrons of the CP are the same (triplet injection).

It is necessary to modify the rates in order to encode the effects of the magnetic flux $\phi$ piercing the junction perpendicularly and of an applied bias $V$ across the two SCs. The former introduces an Aharonov-Bohm (AB) phase in the processes involving a direct Andreev reflection, while the latter brings a time dependence into the superconducting phase difference. We neglect the Zeeman coupling since it only provides a small energy splitting of the energy bands while not significantly contributing to our results. The orbital effect is encoded by the minimal coupling $-i\hbar\nabla\rightarrow -i\hbar\nabla+e \mathbf{A}$, with $e>0$ the absolute value of the electron charge. We use the Landau gauge to set $\mathbf{A}=B(-y,0,0)$, so that the states on the upper and lower edge acquire opposite contributions $\mp \frac{e v_F\phi}{2L}$ to the energy. In the long junction limit $L\gg (\hbar v_F)/\mu$, the orbital contribution can hence also be neglected.

The gauge-invariant phase difference experienced by a CP going from the left to right has the generic form $\varphi_r-\varphi_l=(\varphi^0_r-\varphi^0_l)+\gamma^{AB}+\omega_Jt$, where $\omega_J=2eV/\hbar$ is the Josephson frequency and $\gamma^{AB}$ takes into account the AB phase picked by the two electrons of the CP.

It is well known \cite{loss,flux} that a single electron travelling (from left to right) on the upper/lower edge all along the junction acquires an AB phase $\gamma^{AB}_e=\pm\pi\phi/2\phi_0$, with $\phi_0=h/2e$ the superconducting flux quantum. Therefore, in the absence of inter-edge tunnelling, $\gamma^{AB}=\pm\pi\phi/\phi_0$ for a CP that enters the 2DTI via a DAR process and exits with another DAR process on the upper/lower edge. By contrast, if a CP is injected and extracted via two CAR processes, its AB phase is $\gamma^{AB}=0$, since the AB phases picked by each electron cancels with each other.

The intriguing peculiarity of our system, however, is that we allow for an arbitrary number of inter-edge tunnellings in the constriction, and this allows for novel processes that transfer a CP between the SCs. To characterise these possibilities, it is important to distinguish between even and odd numbers of tunneling events.\\
With even number of tunnellings, a CP can now enter the junction with a DAR process on the upper/lower edge and leave it with a DAR process on the opposite (i.e. lower/upper) one. In this case, the AB phase is zero, since the phase picked by the two electron on the upper edge is cancelled by the one picked on the lower one.\\
With an odd number of tunnellings, by contrast, a CP can enter the junction with a DAR and exit with a CAR (or viceversa). In this scenario, the unconventional AB phase picked by the CP is given by $\gamma^{AB}=\pm\pi\phi/2\phi_0$. Let us clarify this point with an example, considering a CP that enters the junction from the left with a DAR in the upper edge and exits it to the right with a CAR process. In this case, the two electrons pick the same phase during their transit on the upper edge, from the left SC to the constriction. However, after the constriction, they travel on opposite edges and the phases that they pick between the constriction and the right SC cancel with each other. As a result, $\gamma^{AB}=\pi\phi/2\phi_0$. The opposite sign is obtained when the DAR process happens on the lower edge.

In deriving the expressions of the AB phases above, we neglected the flux enclosed in the constriction. This approximation, which is fully justified by the assumption $\ell w\ll WL$, allows us not to care about the exact number of tunnelling events and their exact location within the constriction. If the flux enclosed in the constriction were included in our calculation, it would just add a weak Fraunhofer-like decay to the flux-dependence, but our results would be qualitatively unchanged.

Interestingly, we can include in our previous description all the AB phases described before, just by adding flux-dependent factors to the tunnelling amplitudes in Eq. (\ref{eqn:gammaloss}):
\begin{equation}
e^{-i\varphi^0_j}\rightarrow e^{-i\left[\varphi^0_j+j\frac{1}{2}\left(\omega_Jt+\frac{\pi\phi(\tau_1+\tau_2)}{2\phi_0}\right)\right]}.\label{eqn:phase}
\end{equation}
Notice that the gauge-invariant phase difference mentioned above is among the SCs, and hence involves one tunnelling with $j=-1$ ($\Gamma^*_{\zeta_1,\zeta_2,-1}$, from the left SC to the edges) and one with $j=1$ ($\Gamma_{\zeta_3,\zeta_4,1}$, from the edges to the right SC). Eq. (\ref{eqn:phase}) returns the correct phase differences we discussed: for two CAR processes ($\tau_1=-\tau_2$), there is no flux dependence; for two DAR processes on the same edge $\tau$, we do get a flux dependent phase $\tau\pi\phi/\phi_0$; for two DAR on opposite edges we get zero; for a CAR and a DAR we get the anomalous term, with the sign depending on the $\tau$ of the DAR.

Finally, we make the unitary transformation inverse to the one of Eq. (\ref{eqn:En.A}). Let $a_{\zeta_i,i}$, with $\zeta_i=11, \,-11,\,-1-1,\,1-1$ and $i=1,\,2,\,3,\,4$, be the elements of such unitary matrix. Then, for each of the four operators $\hat{c}_{k,\zeta_i}$, we have
	\begin{equation}
		\hat{c}_{k,\zeta_i}=a_{\zeta_i,1}\hat{A}_{k,1}+a_{\zeta_i,2}\hat{A}_{k,2}+a_{\zeta_i,3}\hat{A}_{k,3}+a_{\zeta_i,4}\hat{A}_{k,4}.\label{eqn:cambiobase}
	\end{equation}
	Substituting Eq. (\ref{eqn:cambiobase}) in Eq. (\ref{eqn:dHk}), we obtain
	\begin{align}
		\delta \hat{H}_E&=\sum_{i_1,i_2}\sum_{k_1,k_2}\sum_{\zeta_1,\zeta_2,j}\frac{1}{L}\left[\Gamma_{\zeta_1,\zeta_2,j}a_{\zeta_1,i_1}a_{\zeta_2,i_2}e^{ik_1x_j^-}e^{ik_2 x_j^+}\hat{A}_{k_1,i_1}\hat{A}_{k_2,i_2}+h.c.\right]\notag\\
		&\equiv\sum_{i_1,i_2,j}\sum_{k_1,k_2}\frac{1}{L}\left[\Gamma_{i_1,i_2,j}(k_1,k_2)\hat{A}_{k_1,i_1}\hat{A}_{k_2,i_2}+h.c.\right],
	\end{align}
	where we have defined
	\begin{equation}
		\Gamma_{i_1,i_2,j}(k_1,k_2)\equiv\sum_{\zeta_1,\zeta_2}\Gamma_{\zeta_1,\zeta_2,j}a_{\zeta_1,i_1}a_{\zeta_2,i_2}e^{ik_1x_j^-}e^{ik_2 x_j^+}.
	\end{equation}
For now, the summation over $i_1,i_2$ runs over 16 terms. We can reduce them up to 10 by antisymmetrising the coefficients when $i_1\neq i_2$, introducing the new coefficients $\alpha_{i_1,i_2}$\footnote{E.g., for $i_1,i_2=1,2$:\\
		\begin{align*}
			&\sum_{k_1,k_2}\Gamma_{1,2,j}(k_1,k_2)\hat{A}_{k_1,1}\hat{A}_{k_2,2}+\sum_{k_1,k_2}\Gamma_{2,1,j}(k_1,k_2)\hat{A}_{k_1,2}\hat{A}_{k_2,1}=\notag\\
			=&\sum_{k_1,k_2}\Gamma_{1,2,j}(k_1,k_2)\hat{A}_{k_1,1}\hat{A}_{k_2,2}+\sum_{k_1,k_2}\Gamma_{2,1,j}(k_2,k_1)\hat{A}_{k_2,2}\hat{A}_{k_1,1}=\notag\\
			=&\sum_{k_1,k_2}\Gamma_{1,2,j}(k_1,k_2)\hat{A}_{k_1,1}\hat{A}_{k_2,2}-\sum_{k_1,k_2}\Gamma_{2,1,j}(k_2,k_1)\hat{A}_{k_1,1}\hat{A}_{k_2,2}=\notag\\
			=&\sum_{k_1,k_2}\left(\Gamma_{1,2,j}(k_1,k_2)-\Gamma_{2,1,j}(k_2,k_1)\right)\hat{A}_{k_1,1}\hat{A}_{k_2,2}\equiv\notag\\
			\equiv&\sum_{k_1,k_2}\alpha_{1,2,j}(k_1,k_2)\hat{A}_{k_1,1}\hat{A}_{k_2,2}.
	\end{align*}}. We obtain
	\begin{equation}
		\delta \hat{H}_E=\sum_j\delta \hat{H}_E^j=\sum_j\sum_{i_1,i_2}\sum_{k_1,k_2}\frac{1}{L}\left[\alpha_{i_1,i_2,j}(k_1,k_2)\hat{A}_{k_1,i_1}\hat{A}_{k_2,i_2}+h.c.\right],\label{eqn:dHfin}
	\end{equation}
where now $\alpha_{i_1,i_2}=\alpha_{1,1}\quad \alpha_{1,2}\quad \alpha_{1,3}\quad \alpha_{1,4}\quad \alpha_{2,2}\quad \alpha_{2,3}\quad \alpha_{2,4}\quad \alpha_{3,3}\quad \alpha_{3,4}\quad \alpha_{4,4}$.

We hence have the effective Hamiltonian $\hat{H}$ of the edges, which reads
\begin{equation}
    \hat{H}=\hat{H}_E^0+\delta \hat{H}_E.
\end{equation}

In the next section, we briefly set up the formalism we use for the evaluation of the Josephson current.

\section{Formalism for the transport properties}
\label{sec3}
Let $\hat{N}=\sum_i\sum_k\hat{A}^{\dagger}_{ki}\hat{A}_{ki}$ be the total number operator relative to the electrons. The net change in the number of electrons, different from zero due to the coupling to the SCs, is given by $\dot{\hat{N}}=\dot{\hat{N}}^r+\dot{\hat{N}}^l$, with
\begin{align}
\dot{\hat{N}}^r&=\frac{i}{\hbar}[\hat{H}_E^0+\delta \hat{H}_E^r,\hat{N}]=\frac{i}{\hbar}[\delta \hat{H}_E^r,\hat{N}]\notag\\
\dot{\hat{N}}^l&=\frac{i}{\hbar}[\hat{H}_E^0+\delta \hat{H}_E^l,\hat{N}]=\frac{i}{\hbar}[\delta \hat{H}_E^l,\hat{N}]
\end{align}
in the Heisenberg picture.

\begin{figure}[h!]
	\centering
	\includegraphics[width=9cm]{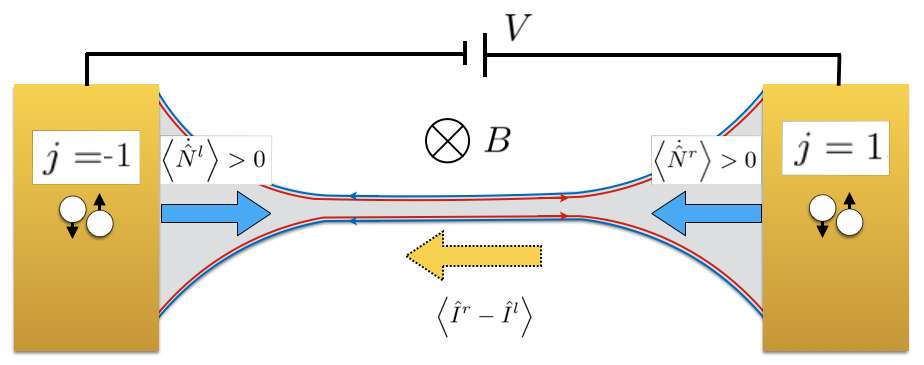}
	\caption{\small{Direction of the right, left and total current flow according to our conventions.}}
	\label{fig:current}
\end{figure}

The operator $\hat{I}^j$, relative to the current flowing in the edges reads as\footnote{We remind that the perturbation $\delta \hat{H}_E^j$ - and therefore also $\hat{I}^j$ - acquires a time dependence in the $\Gamma_{\zeta_1,\zeta_2,j}$ coefficients as the bias $V$ is non-zero (see after Eq. (\ref{eqn:dHk}) and Appendix A).} $\hat{I}^j(t)=e\dot{\hat{N}}^j$. According to this convention, $\hat{I}^r(t)$ and $\hat{I}^l(t)$ are the currents injected from the superconducting leads in the edges, as shown in Fig. \ref{fig:current} by the blue arrows. The total current is
\begin{equation}
\hat{I}^{tot}(t)=\hat{I}^r(t)-\hat{I}^l(t),
\end{equation}
flowing in the direction shown in the figure (yellow arrow).\\
Since the procedure is independent of the choice of the lead, in the following we sketch the calculation of the generic $j^{th}$ term.

Carrying out the anticommutation leads to
	\begin{equation}
		\hat{I}^{j}=\frac{2ie}{\hbar L}\sum_{i_1,i_2}\sum_{k_1,k_2}\frac{1}{L}\left[\alpha_{i_1,i_2,j}(k_1,k_2)\hat{A}_{k_1,i_1}\hat{A}_{k_2,i_2}-h.c.\right].
	\end{equation}
	
Assuming the coupling between the lead $j$ and the edges to be weak (namely $\Gamma N_E\ll1$, with $N_E$ the density of states at the Fermi-level in the edge system), $\delta\hat{H}_E$ can be regarded as a small perturbation. We want to compute the expectation value
\begin{equation}
{I}^{j}(t)\equiv\left\langle \hat{U}(-\infty,t)\hat{I}^{j}(t)\hat{U}(t,-\infty)\right\rangle,
\label{eqn:average}
\end{equation}
taken with respect to the unperturbed edge state system in the far past. Here, $\hat{U}(-\infty,t)=\hat{U}^{\dagger}(t,+\infty)=\hat{T}_+e^{-\frac{i}{\hbar}\int_{-\infty}^t d\tau \delta \hat{H}_E(\tau)}$ is the time-evolution operator in the interaction picture representation, with $\hat{T}_+$ the time-ordering operator.

According to the linear response theory \cite{vignale}, that is to say up to linear order, in $\delta\hat{H}_E$ (second order in $\Gamma$ in this case), we get
\begin{equation}
{I}^{j}(t)\approx\frac{i}{\hbar}\int_{-\infty}^t d\tau\left\langle\left[\delta \hat{H}_E(\tau),\hat{I}^{j}(t)\right]\right\rangle_{0}.\label{eqn:linresp}
\end{equation}
The subscript ``0" serves as a reminder of the fact that we are considering an equilibrium average calculated with respect to the unperturbed system (described by $\hat{H}_E^0$, see Eq. (\ref{eqn:H_E^0A})) in the past. In the calculation of $I^j(t)$, the relevant perturbation is the one induced by the $-j^{th}$ superconductor, namely $\delta \hat{H}^{-j}_E$. This is easily understood by looking at the expressions of the $\Gamma_{\zeta_1,\zeta_2,j}$ coefficients: since they keep trace of the phase of the superconductor to whom they are related, the SCs phase difference-dependent supercurrent originates from the $-j,\,j$ terms.

Moving forward with the calculations (see Appendix B for the details), we get
	\begin{align}
		{I}^{j}(t)=&\frac{8e}{\hbar^2}\text{Im}\bigg\{\int_{-\infty}^{\infty}dt'\theta(t')\sum_{i_1,i_2}\sum_{k_1,k_2,k_1',k_2'}\frac{1}{L^2}\alpha_{i_1,i_2,j}(k_1',k_2',t)\alpha^*_{i_1,i_2,-j}(k_1,k_2,t-t')\notag\\
		&\left\langle\left[\hat{A}_{k_1',i_1}(t)\hat{A}_{k_2',i_2}(t),\hat{A}^{\dagger}_{k_2,i_2}(t-t')\hat{A}^{\dagger}_{k_1,i_1}(t-t')\right]\right\rangle_{0}\bigg\},\label{eqn:current}
	\end{align}
where we have recovered the explicit time-dependence of the coefficients $\alpha$ and the correlation functions from now on will be time-ordered.

The unperturbed edge system being time-translation invariant, Eq. (\ref{eqn:current}) is equivalent to
	\begin{align}
		{I}^{j}(t)=&\frac{8e}{\hbar^2}\text{Im}\bigg\{\int_{-\infty}^{\infty}dt'\theta(t')\sum_{i_1,i_2}\sum_{k_1,k_2,k_1',k_2'}\frac{1}{L^2}\alpha_{i_1,i_2,j}(k_1',k_2',t)\alpha^*_{i_1,i_2,-j}(k_1,k_2,t-t')\notag\\
		&\left\langle\left[\hat{A}_{k_1',i_1}(t')\hat{A}_{k_2',i_2}(t'),\hat{A}^{\dagger}_{k_2,i_2}(0)\hat{A}^{\dagger}_{k_1,i_1}(0)\right]\right\rangle_{0}\bigg\},\label{eqn:current2}
	\end{align}
where the time-evolved operators in the expectation value are simply given by
	\begin{equation}
		\hat{A}_{k_1',i_1}(t')=\hat{A}_{k_1',i_1}e^{-iE_{A_{i_1}}(k_1')t'/\hbar},\,\hat{A}_{k_2',i_2}(t')=\hat{A}_{k_2',i_2}e^{-iE_{A_{i_2}}(k_2')t'/\hbar}.
	\end{equation}
	
Making use of Wick's theorem and recalling that
	\begin{equation}
		\left\langle\hat{A}_{k_1,i_1}(t')\hat{A}^{\dagger}_{k_2,i_2}(0)\right\rangle_{0}=\delta_{i_1,i_2}\delta_{k_1,k_2}\frac{1}{1+e^{\beta E_{A_{i_1}}(k)}},
	\end{equation}
	with $\beta=1/k_BT$, we obtain
	\small
	\begin{align}
		\sum_{k_1,k_2,k_1',k_2'}&\frac{1}{L^2}\alpha_{i_1,i_2,j}(k_1',k_2',t)\alpha^*_{i_1,i_2,-j}(k_1,k_2,t-t')\left\langle\left[\hat{A}_{k_1',i_1}(t')\hat{A}_{k_2',i_2}(t'),\hat{A}^{\dagger}_{k_2,i_2}(0)\hat{A}^{\dagger}_{k_1,i_1}(0)\right]\right\rangle_{0}=\notag\\
		=\frac{1}{(2\pi)^2}\bigg\{-&\delta_{i_1,i_2}\int_{-\infty}^{+\infty} dk_1\,e^{-iE_{A_{i_1}}(k_1)t'/\hbar}\frac{e^{\beta E_{A_{i_1}}(k_1)}}{1+e^{\beta E_{A_{i_1}}(k_1)}}\cdot\notag\\
		\cdot&\int_{-\infty}^{+\infty} dk_2\,\alpha_{i_1,i_1,j}(k_1,k_2,t)\alpha^*_{i_1,i_1,-j}(k_1,k_2,t-t')e^{-iE_{A_{i_1}}(k_2)t'/\hbar}\frac{e^{\beta E_{A_{i_1}}(k_2)}}{1+e^{\beta E_{A_{i_1}}(k_2)}}+\notag\\
		+&\int_{-\infty}^{+\infty} dk_1\,e^{-iE_{A_{i_1}}(k_1)t'/\hbar}\frac{e^{\beta E_{A_{i_1}}(k_1)}}{1+e^{\beta E_{A_{i_1}}(k_1)}}\cdot\notag\\
		\cdot&\int_{-\infty}^{+\infty} dk_2\, \alpha_{i_1,i_2,j}(k_1,k_2,t)\alpha^*_{i_1,i_2,-j}(k_1,k_2,t-t')e^{-iE_{A_{i_2}}(k_2)t'/\hbar}\frac{e^{\beta E_{A_{i_2}}(k_2)}}{1+e^{\beta E_{A_{i_2}}(k_2)}}\bigg\}.\label{eqn:current3}
	\end{align}
\normalsize
The full expansion of Eq. (\ref{eqn:current2}) is cumbersome. However, Eqs. (\ref{eqn:current2})-(\ref{eqn:current3}) are useful to identify the typical structure of each term: a product of two $\alpha$ coefficients - which, through the $\Gamma$s, contain all the details specifying the tunnelling process - and two Green functions given by the two integrals.

\section{Results}
\label{sec4}
\subsection{Analytical results}
The long and cumbersome calculation sketched in the previous section leads to the following expression for the right/left current
	\begin{align}
		I^{r/l}(t)&=\mathcal{C}\,\text{Im}\bigg\{e^{\mp i(\omega_J t+\varphi^0_r-\varphi^0_l)}\int_0^{+\infty}ds\,e^{\pm is\tilde{V}}\left[A_1\cos{\left(\pi\frac{\phi}{\phi_0}\right)}+A_2\sin{\left(\frac{\pi}{2}\frac{\phi}{\phi_0}\right)}+A_3\right]\bigg\}=\notag\\
		&=\mathcal{C}\,\text{Im}\bigg\{e^{\mp i(\omega_J t+\varphi^0_r-\varphi^0_l)}\left[\tilde{A}^{r/l}_1\cos{\left(\pi\frac{\phi}{\phi_0}\right)}+\tilde{A}_2^{r/l}\sin{\left(\frac{\pi}{2}\frac{\phi}{\phi_0}\right)}+\tilde{A}_3^{r/l}\right]\bigg\}.\label{eqn:currentflux}
	\end{align}
Here, $\mathcal{C}\equiv(-2e\Delta\Gamma^2)/(\pi^2\hbar^3v_F^2)$ is a constant. The dimensionless quantities $\tilde{V}=eV/\Delta$ and $s=\frac{t'\Delta}{\hbar}$, with $t'$ a time variable have been introduced. Notice that $A_1,\,A_2$ and $A_3$ are three complex factors that depend on all the parameters except for the flux $\phi$. Likewise, $\tilde{A}^{r/l}_1,\,\tilde{A}^{r/l}_2,\,\tilde{A}^{r/l}_3$ stand for the three coefficients $A_1,\,A_2,\,A_3$ once the integration over $s$ is done. More details can be found in Appendix B.

We draw the attention to the term proportional to $\sin{\left(\frac{\pi}{2}\frac{\phi}{\phi_0}\right)}$. Indeed, it encodes a $4\pi$-periodicity with respect to the magnetic flux. This term is absent in the absence of forward scattering among the two edges ($f=0$). Moreover, in order to have a nonzero $\tilde{A}^{r/l}_2$ one must additionally have $\tilde{f}_T\neq 0$ provided by the strong spin-orbit coupling characterising the structure. This is due to the fact that an odd number of tunnellings within the constriction necessarily leads to a spin-flip of one electron, that has to be counterbalanced by an additional spin-flip at the SC-2DTI interface. What the analytical calculation shows is that the physical processes related to the anomalous periodicity are the ones in which the CP switches from being on one edge only to being delocalised between the two edges. These processes are proportional to $\tilde{f}_Tf_C$ and must include one forward tunnelling in the constriction (or an odd number of them).

\begin{figure}
		\centering
		\includegraphics[width=9cm]{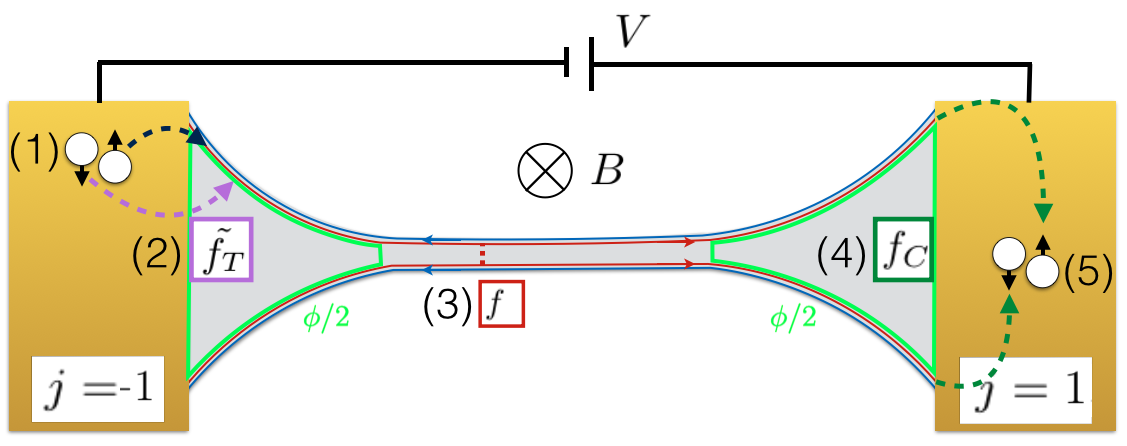}
		\caption{\small{A prototypical example of a process carrying a $4\pi$-periodic dependence on the flux $\phi$: (1) a CP is injected from the left SC in the upper edge; (2) during the injection, one of the electrons is spin-flipped, and thus the CP enters the edge system in a spin-triplet state, acquiring a $\tilde{f}_T$ factor accounting for the spin-flip. (3) Having enclosed a flux $\phi/2$, in the constriction an electron tunnels to the lower edge; (4) when the opposite end of the junction is reached, a final $f_C$ factor keeps track of the CAR; (5) clearly, the CP reaches the right SC in a spin-singlet state.}}
		\label{fig:4pi}
	\end{figure}

An example is depicted in Fig. \ref{fig:4pi}, where a CP enters the junction with a DAR in the upper edge and exits it with a CAR after one electron tunnels in the constriction. The AB phases picked up by the two electrons sum up on the left side of the constriction and cancel on the right side, while no meaningful phase is picked along the constriction (making the precise location of the tunnelling within the constriction irrelevant). The amplitude of the process just described is proportional to $\Gamma_{11,1-1,1}\Gamma^*_{11,11,-1}$, which gives
\begin{align*}
    \Gamma_{11,1-1,1}\Gamma^*_{11,11,-1}&=\left(\Gamma f_C\,e^{-i\left(\varphi^0_r+\frac{1}{2}\omega_Jt\right)}\right)\left(\Gamma\tilde{f}_T\,e^{i\left(\varphi^0_l-\frac{1}{2}\omega_Jt-\frac{\pi\phi}{2\phi_0}\right)}\right)=\\
    &=\Gamma^2 f_C\tilde{f}_T e^{-i\left[\left(\varphi^0_r-\varphi^0_l\right)+\omega_Jt\right]}e^{-i\frac{\pi\phi}{2\phi_0}}.
\end{align*}
Notice that the flux dependence does correspond to what we commented before. A spin-flip $\tilde{f}_T$ is needed due to the fact that in (1) and (5) in the figure we have a singlet state and that within the constriction an odd number (one) of inter-edge tunnellings results in a spin-flip of one electron.

The anomalous periodicity hence represents a hallmark of the presence of forward scattering and the coupling between the edges. Note that it is in principle difficult to demonstrate the presence of such a coupling. Indeed, as already discussed, in long constrictions the single particle backscattering is only able to open a gap at the Dirac point. However such point is often hidden in the valence band of the structure \cite{nat,bands,bands2}. Achieving zero conductance as a function of the gate voltage, that would strongly indicate the formation of the constriction, is hence not always possible and was never experimentally realised. At present, the signature that is related to the existence of the constriction is a reduction of the conductance from $2e^2/h$ to $e^2/h$. However, this behaviour is due to electronic interactions and is hence unclear if it would persist in the presence of superconductors in close contact to the constriction. A clearer signature on the formation of the constriction is thus in order.

\subsection{Quantitative analysis}
In this subsection, in addition to the already introduced $\tilde{V}$, we define the natural dimensionless quantities
\begin{align*}
		\tilde{L}=\frac{L\Delta}{\hbar v_F},\qquad \tilde{T}=\frac{\pi k_B T}{\Delta},\qquad\tilde{\mu}=\frac{\mu}{\Delta},\qquad \tilde{f}=\frac{f}{\Delta}
\end{align*}
to better understand the relevant scales. Notice that our previous assumption $E\ll\Delta$ requires $\tilde{V},\,\tilde{T}\ll1$.

\begin{figure}
		\centering
		\includegraphics[width=10cm]{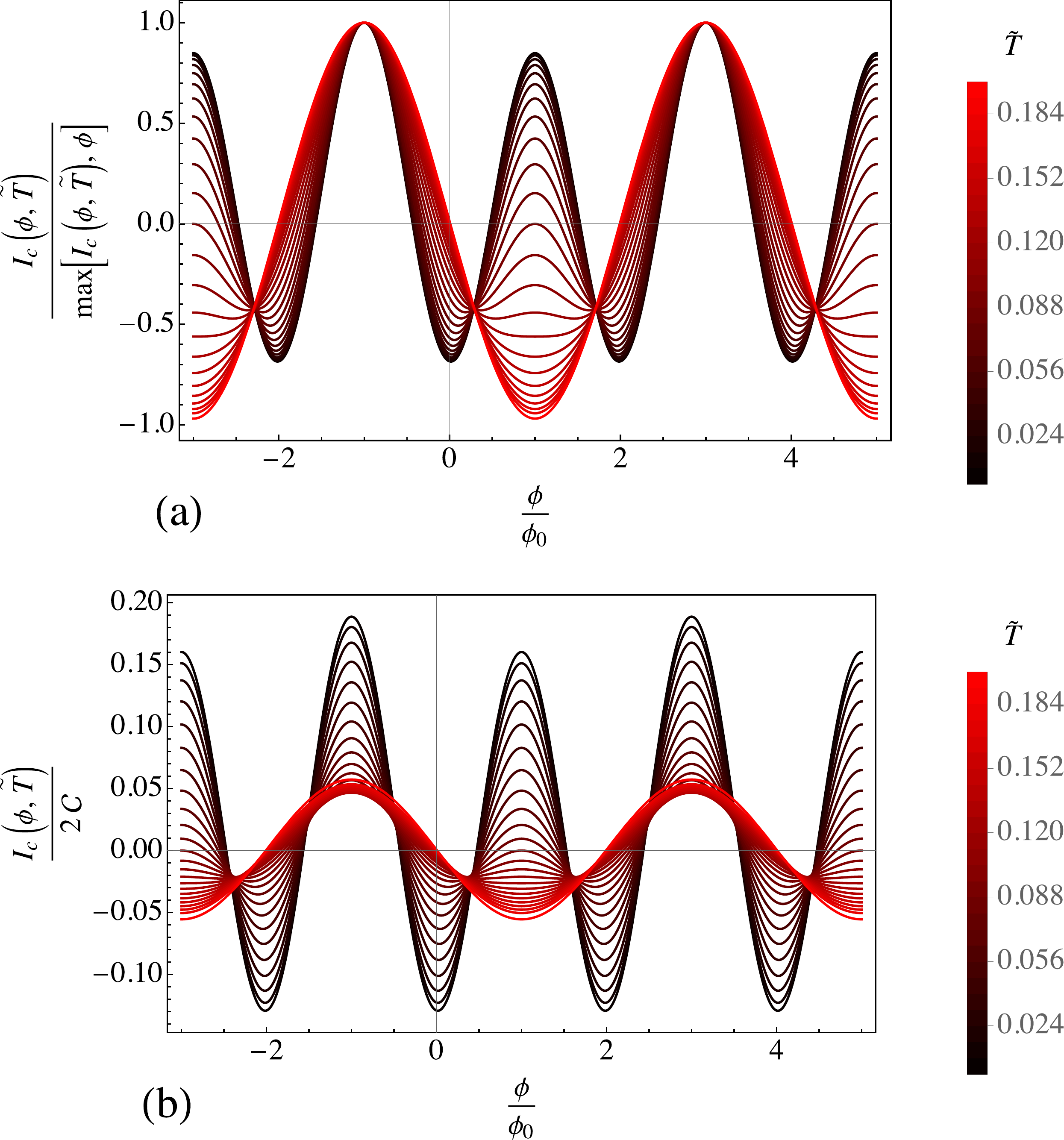}
		\caption{\small{(a) $I_{c}$, normalised with respect to its maximum, for $\tilde{L}=20$, $k_FL=6\pi$, $\tilde{f}_T=0.4$, $f_C=0.3$, $\tilde{\mu}=10^{-4}$ and $\tilde{f}=0.3$. The plot is represented as a function of $\phi$ (in units of $\phi_0$) and for $\tilde{V}=0$. We have $\tilde{T}$ varying between 0.008 and 0.2 (see the red bar legend aside).\\
		(b) Same but without the normalisation to the maximum.}}
		\label{fig:dc}
	\end{figure}

In the limit $\tilde{V}\rightarrow 0$, we inspect the critical current $I_c$, given by the total current $I^r-I^l$ for $\varphi^0_r-\varphi^0_l=\pi/2$. The anomalous periodicity is more pronounced at high temperature $\tilde{T}$, while at low temperature, a $2\pi$-periodicity is almost recovered, as shown in Fig. \ref{fig:dc}(a). There, the current at different temperatures, normalised to its maximum value, is shown as a function of the flux. This scaling allows to easily compare the periodicity of the curves. In the picture, darker lines correspond to lower temperatures. However, the magnitude of the current itself decreases as temperature is raised, as shown in Fig. \ref{fig:dc}(b), where the plots are not scaled.

\begin{figure}[h!]
		\centering
		\includegraphics[width=10cm]{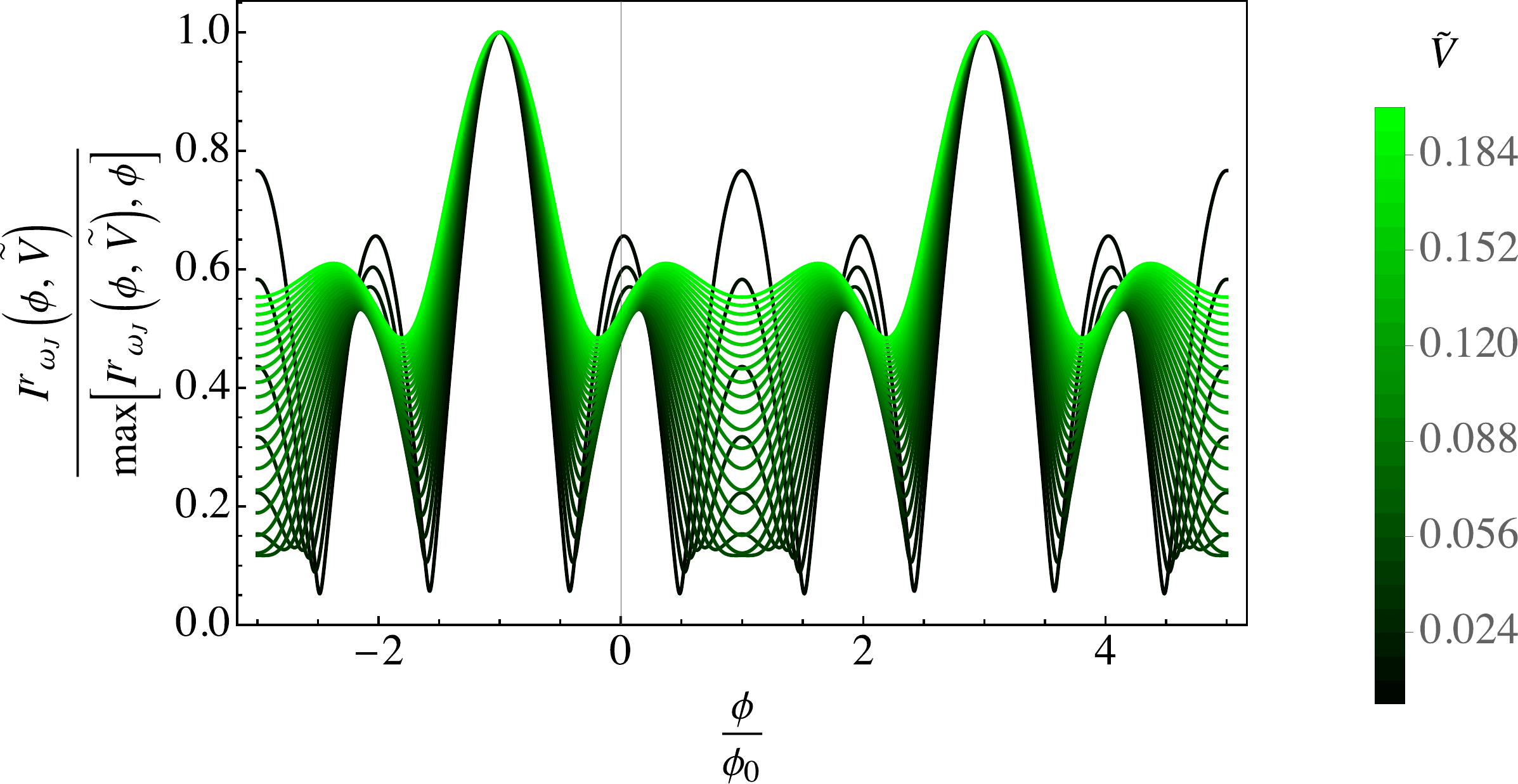}
		\caption{\small{$I^r_{\omega_J}$, normalised with respect to its maximum, for $\tilde{L}=20$, $k_FL=6\pi$, $\tilde{f}_T=0.4$, $f_C=0.3$, $\tilde{\mu}=10^{-4}$, $\tilde{f}=0.3$ and $\tilde{T}=0.1$. The plot is represented as a function of $\phi$ (in units of $\phi_0$) and for $\tilde{V}$ varying between 0.008 and 0.2 (see the green bar legend aside).}}
		\label{fig:V}
	\end{figure}

\begin{figure}
		\centering
		\includegraphics[width=10cm]{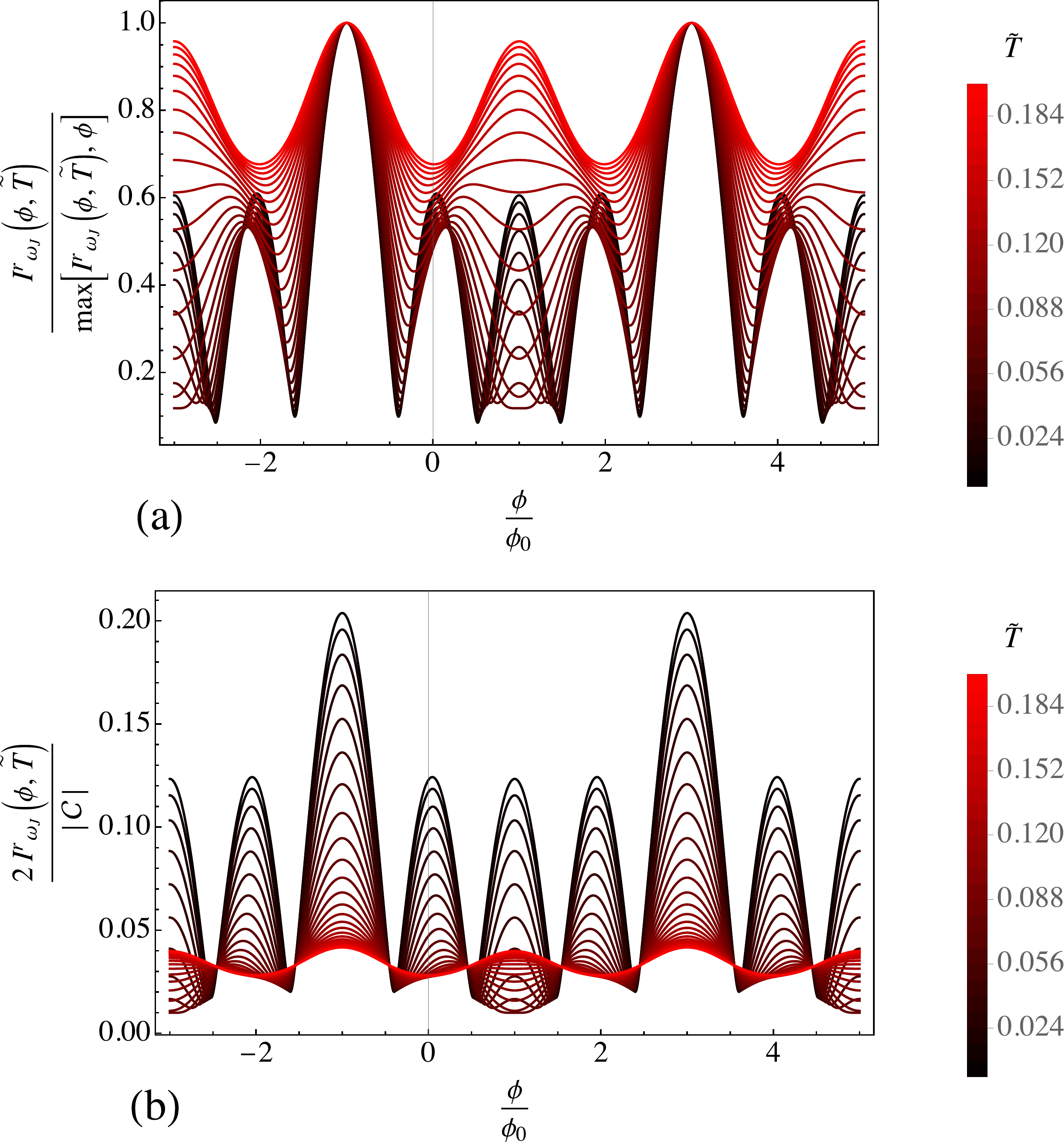}
		\caption{\small{(a) $I^r_{\omega_J}$, normalised with respect to its maximum, for $\tilde{L}=20$, $k_FL=6\pi$, $\tilde{f}_T=0.4$, $f_C=0.3$, $\tilde{\mu}=10^{-4}$, $\tilde{f}=0.3$ and $\tilde{V}=0.1$. The plot is represented as a function of $\phi$ (in units of $\phi_0$) and for $\tilde{T}$ varying between 0.008 and 0.2 (see the red bar legend aside).\\
		(b) Same but without the normalisation to the maximum.}}
		\label{fig:VT}
\end{figure}

In the finite $\tilde{V}$ regime the current becomes time dependent, in view of the AC Josephson effect. As in Ref. \cite{loss}, we hence analyse the Fourier component of the current at the frequency $\omega_J$ given by the voltage. Quantitatively, what we analyse is 
\begin{equation}
I_{\omega_J}^r=\left|\frac{1}{\text{T}}\int_{-\text{T}/2}^{\text{T}/2}e^{-i\omega_Jt}I^r(t)\right|,
\end{equation}
with $\text{T}=2\pi/\omega_J$. The Fourier transform of the total current is qualitatively similar. We observe that the applied bias increases the visibility of the anomalous periodicity, as shown in Fig. \ref{fig:V}. On top of that, and differently from the zero bias case, the $4\pi$ component becomes more pronounced for lower temperature (Fig. \ref{fig:VT}(a)). Moreover, as in the zero bias limit, higher temperature makes the signal smaller. This fact is shown in Fig. \ref{fig:VT}(b), which represents the same plots as Fig. \ref{fig:VT}(a) but without scaling. The zero bias and the finite bias regime, exhibiting different behaviours to varying of the parameters, can hence be useful for different materials. 

\section{Discussion and conclusions}
\label{sec5}
In this work, we have analysed a Josephson junction where the non-superconducting element is a long constriction between helical edge states, pierced by a magnetic flux. We have shown that the interplay between the strong spin-orbit interaction characterising the system and the coupling between the edges results in an anomalous $4\pi$-periodicity of the Josephson current with respect to the flux. This behaviour is present both in the DC and in the AC Josephson effect. It is robust with respect to temperature and no fine tuning is necessary. The effect in our system needs the presence of three properties: a DAR with spin-flipping, a tunnelling between the edges and a CAR, as shown in Fig. \ref{fig:4pi}. Throughout our discussion, we neglected the magnetic flux enclosed in the constriction; if included, it would just add a weak Fraunhofer-like decay to the pattern in our plots, without affecting significantly our results.

The anomalous periodicity we found represents a hallmark of a constriction between edges tunnel-coupled to superconductors. Its physical origin resides in the possibility of switching the CP nature (localised on one edge/delocalised on both edges) enabled by the single-electron tunnelling.

\begin{figure}[h!]
		\centering
		\includegraphics[width=10cm]{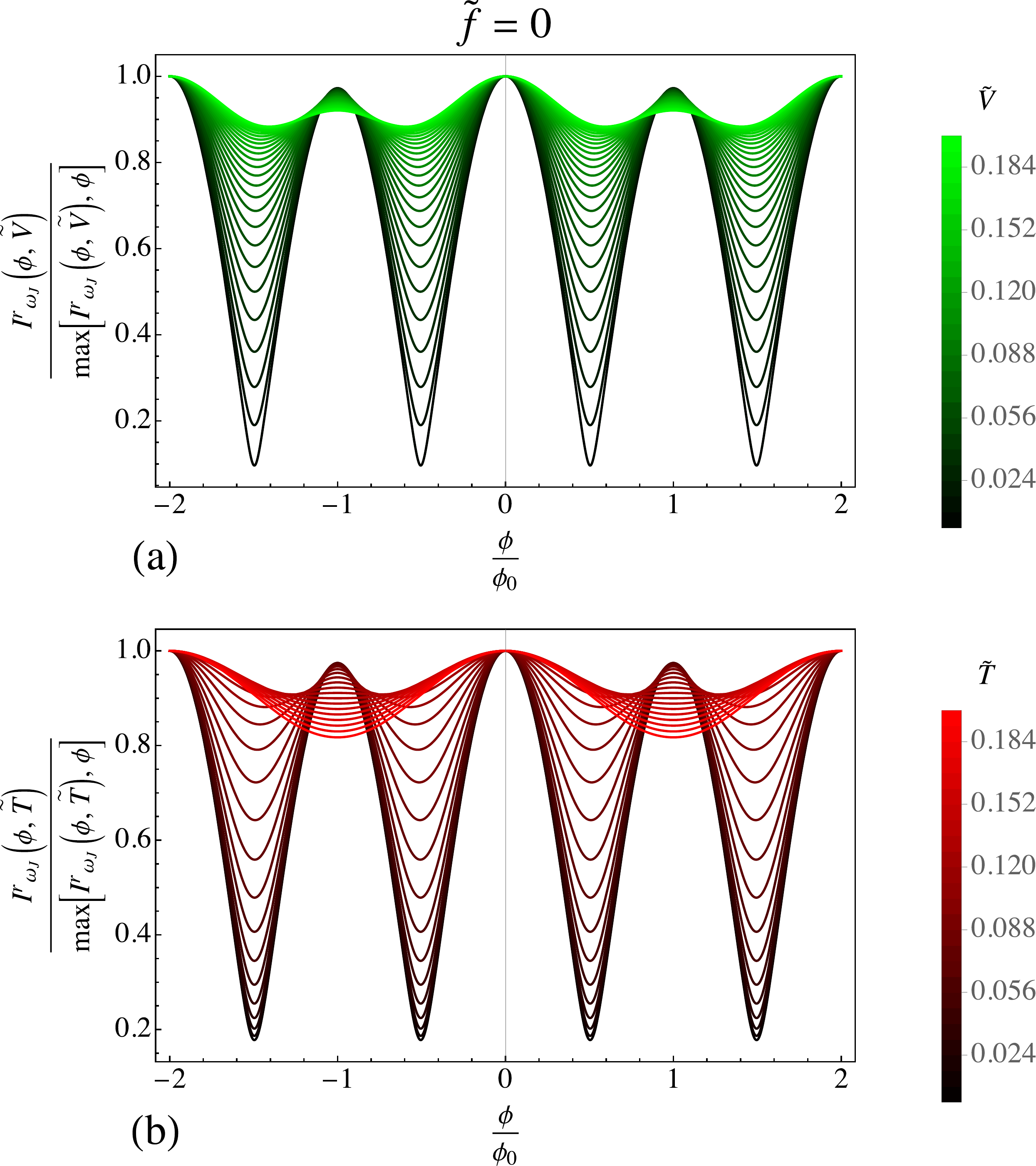}
		\caption{\small{(a) $I^r_{\omega_J}$, normalised with respect to its maximum, for $\tilde{L}=20$, $k_FL=6\pi$, $\tilde{f}_T=0.4$, $f_C=0.3$, $\tilde{\mu}=10^{-4}$, $\tilde{f}=0$ and $\tilde{T}=0.1$. The plot is represented as a function of $\phi$ (in units of $\phi_0$) and for $\tilde{V}$ varying between 0.008 and 0.2 (see the green bar legend aside).\\
		(b) Same parameters values, but now $\tilde{V}=0.1$ and $\tilde{T}$ varies between 0.008 and 0.2 (see the red bar legend aside).}}
		\label{fig:2pi}
\end{figure}

In the absence of inter-edge tunnelling, the fate of the CPs is known from the very beginning: CPs entering the weak link in DAR on the left end ($x=-L/2,\,y=\pm W/2$) necessarily end up on the right end of the same edge ($x=L/2,\,y=\pm W/2$), picking a phase difference $\pm \pi\phi/\phi_0$; analogously, CPs entering the link in CAR unavoidably leave it in CAR without collecting any phase, as discussed in Subsec. (\ref{subsec:2.2}). This scenario corresponds to a standard SQUID pattern, similar to \cite{loss}, and is plotted in Fig. \ref{fig:2pi}.\\
On the other hand, the inter-edge tunnelling we included broadens the possibilities, in particular allowing a CP to enter the junction in DAR and leave it in CAR or viceversa. The single-electron tunnelling marks the boundary between two kinds of CP, with different insights of the magnetic flux. For the setup we sketched, the phase difference in this case amounts to $\pm\pi\phi/2\phi_0$, leading to a $4\pi$-periodic contribution to the Josephson current.\\
To summarise our finding: if $f=0$ (Fig. \ref{fig:2pi}), there are a $\pi$ and a $2\pi$ components (the latter being usually known as even-odd effect, due to the presence of CAR, as discussed for instance in \cite{loss,evenodd}); temperature and bias allow to make the $2\pi$ term more or less prominent (as commented in \cite{loss}). With a finite $f$ (Figs. \ref{fig:dc}-\ref{fig:VT}), a novel $4\pi$ component arises (coexisting with the other two that are more or less still present).

We argue that, whenever in a constriction there is the possibility of a swap from a CP living on one side to a CP living on both sides, anomalous periodicities can emerge. In this sense, they are more generally a phenomenon related to single electron physics in the superconducting context. Applied to the setup under investigation, this physical interpretation substantiate the robustness of the anomalous periodicity with respect to slight spatial variations of the parameters.

While our discussion is general and applies to all the quantum spin Hall (candidate) systems, we argue that our results are readily observable in HgTe-CdTe heterostructures. Indeed, with the realistic length scales $L\simeq3\,\mu m$, $\ell\simeq1\,\mu m$, $W\simeq200\,nm$ and $w\simeq30\,nm$, the magnetic field necessary to observe the effect is of the order $B\simeq10^{-3}\,T$. The temperature range is limited by the pairing potential $\Delta\sim0.40\,meV$ (a reasonable value for HgTe-based systems \cite{values,hajer}), which determines an upper bound of $\sim300\,mK$. Similarly, the bias should not exceed $\sim80\mu eV$.
We mention in passing that the edge states can be robust even in the presence of magnetic fields that are much stronger than the ones needed for our proposal \cite{mag1,mag2,mag3}.

\subsection*{Acknowledgments}
This work was supported by the ``Dipartimento di Eccellenza MIUR 2018-2022", by the W\"urzburg-Dresden
Cluster of Excellence on Complexity and Topology in
Quantum Matter (EXC2147, project-id 390858490) and
by the DFG (SPP1666 and SFB1170 ``ToCoTronics”).

\section*{Appendix A: Hamiltonian of the SCs and tunnelling Hamiltonian}
The $j^{th}$ superconductor Hamiltonian, written in the usual Nambu basis, is given, in the absence of the magnetic field, by \cite{degennes}
\small
\begin{equation}
\hat{H}_S^j=\frac{1}{2}\int d\mathbf{r} \left(\hat{\Psi}^{\dagger}_{j,\uparrow}(\mathbf{r}),\hat{\Psi}^{\dagger}_{j,\downarrow}(\mathbf{r}),\hat{\Psi}_{j,\uparrow}(\mathbf{r}),\hat{\Psi}_{j,\downarrow}(\mathbf{r})\right)
\mathcal{H}_S^j
\begin{pmatrix}\hat{\Psi}_{j,\uparrow}(\mathbf{r})\\\hat{\Psi}_{j,\downarrow}(\mathbf{r})\\\hat{\Psi}^{\dagger}_{j,\uparrow}(\mathbf{r})\\\hat{\Psi}^{\dagger}_{j,\downarrow}(\mathbf{r})\end{pmatrix},\label{eqn:HSCs}
\end{equation}
\normalsize
with
\begin{equation}
  \mathcal{H}_S^j=  \begin{pmatrix}
-\frac{\hbar^2\nabla^2_{\mathbf{r}}}{2m}-\mu_{SC}&0&0&-\Delta e^{-i\varphi_j^0}\\
0&-\frac{\hbar^2\nabla^2_{\mathbf{r}}}{2m}-\mu_{SC}&\Delta e^{-i\varphi_j^0}&0\\
0&\Delta e^{i\varphi_j^0}&\frac{\hbar^2\nabla^2_{\mathbf{r}}}{2m}+\mu_{SC}&0\\
-\Delta e^{i\varphi_j^0}&0&0&\frac{\hbar^2\nabla^2_{\mathbf{r}}}{2m}+\mu_{SC}
\end{pmatrix}.
\end{equation}
As mentioned in the main text, $\Delta$ is the superconducting pairing  amplitude, $\mu_{SC}$ the chemical  potential - which we assume to be the same in the two SCs, while $\varphi_j^0$ are the bare pairing phases that we keep distinguished in principle. Moreover, $\hat{\Psi}_{j,\uparrow/\downarrow}(\mathbf{r})$ is the annihilation operator of an electron in the $j^{th}$ superconducting lead, with $\uparrow/\downarrow$ spin polarisation (with respect to the same axis as the spin polarisation defined in the edges) and at position $\mathbf{r}$. Eq. (\ref{eqn:HSCs}) is nothing but the Hamiltonian of a standard BCS superconductor.

The coupling between the superconductors and the constriction is modeled by means of the time-reversal invariant Hamiltonian \cite{loss} $\hat{H}_T=\sum_{j}\hat{H}_T^j$, with
\begin{equation}
\hat{H}_T^j=\int dx\int d\mathbf{r} \left(\hat{\Psi}^{\dagger}_{j,\uparrow}(\mathbf{r}),\hat{\Psi}^{\dagger}_{j,\downarrow}(\mathbf{r})\right)\mathcal{T}^j(\mathbf{r},x)\begin{pmatrix}\hat{\psi}_{11}(x)\\\hat{\psi}_{-11}(x)\\\hat{\psi}_{-1-1}(x)\\\hat{\psi}_{1-1}(x)\end{pmatrix}+h.c.,\label{eqn:Htun}
\end{equation}
\normalsize
where
\small
\begin{align}
\mathcal{T}^j(\mathbf{r},x)=&\begin{pmatrix}
e^{ik_Fx}\delta(\mathbf{r}-\mathbf{r}_{j,1})&e^{-ik_Fx}\delta(\mathbf{r}-\mathbf{r}_{j,1})(if_T)&e^{-ik_Fx}\delta(\mathbf{r}-\mathbf{r}_{j,-1})&e^{ik_Fx}\delta(\mathbf{r}-\mathbf{r}_{j,-1})(if_T)\\
e^{ik_Fx}\delta(\mathbf{r}-\mathbf{r}_{j,1})(if_T)&e^{-ik_Fx}\delta(\mathbf{r}-\mathbf{r}_{j,1})&e^{-ik_Fx}\delta(\mathbf{r}-\mathbf{r}_{j,-1})(if_T)&e^{ik_Fx}\delta(\mathbf{r}-\mathbf{r}_{j,-1})
\end{pmatrix}\notag\\
&\frac{\mathfrak{T}}{\sqrt{1+f_T^2}}\delta(x-j\frac{L}{2}).\label{eqn:Tmatrix}
\end{align}
\normalsize
Here $k_F$ is the Fermi momentum in the constriction. Moreover $\mathbf{r}_{j,\tau}=(jL/2,\tau W/2,0)^T$ are the contact points between the $j$-th superconductor and the edges ($\mathbf{r}_{j,1}$ for the upper and $\mathbf{r}_{j,-1}$ for the lower edge channel respectively). In Eq. (\ref{eqn:Tmatrix}) two new parameters have been introduced: $f_T$ and $\mathfrak{T}$. The quantity $f_T$ is the ratio of spin-reversing processes over the spin-conserving ones. It is reasonable to include such a parameter in the model since it allows to take into account the Rashba coupling in the material, which makes spin flips possible. Typically, it is $f_T\ll1$. The other parameter $\mathfrak{T}$ is the tunnelling coefficient related to the opacity of the barrier. These parameters are present in the $\Gamma_{\zeta_1,\zeta_2,j}$ coefficients as discussed in Sec. (\ref{sec2}).
	
\section*{Appendix B: Calculation of the current}
In the following, we address the calculation of the current $I^j(t)$ in more detail. Let's look back at Eq. (\ref{eqn:current3}), and make use of the two definitions in the main text
\begin{align*}
	\alpha_{i_1,i_2,j}(k_1,k_2)&\equiv\Gamma_{i_1,i_2,j}(k_1,k_2)-\Gamma_{i_2,i_1,j}(k_2,k_1)\\
	\Gamma_{i_1,i_2,j}(k_1,k_2)&\equiv\sum_{\zeta_1,\zeta_2}\Gamma_{\zeta_1,\zeta_2,j}a_{\zeta_1,i_1}a_{\zeta_2,i_2}e^{ik_1x_j^-}e^{ik_2 x_j^+},
\end{align*}
where we remind that the possible values for $i_1,i_2$ are
\begin{align}
1,1\qquad1,2\qquad1,3\qquad1,4\qquad2,2\qquad2,3\qquad2,4\qquad3,3\qquad3,4\qquad4,4
\end{align}
and those for $\zeta_1,\zeta_2$ are
\begin{align}
&11,11&\qquad&11,-11&\qquad&11,-1-1&\qquad&11,1-1&\qquad&-11,-11\notag\\
&-11,-1-1&\qquad&-11,1-1&\qquad&-1-1,-1-1&\qquad&-1-1,1-1&\qquad&1-1,1-1.\label{eqn:zeta}
\end{align}

We focus on the calculation of $I^j(t)$ for $j=r=1$. In this case, Eq. (\ref{eqn:current3}) becomes
\footnotesize
\begin{align}
		&\sum_{k_1,k_2,k_1',k_2'}\frac{1}{L^2}\alpha_{i_1,i_2,1}(k_1',k_2',t)\alpha^*_{i_1,i_2,-1}(k_1,k_2,t-t')\left\langle\left[\hat{A}_{k_1',i_1}(t')\hat{A}_{k_2',i_2}(t'),\hat{A}^{\dagger}_{k_2,i_2}(0)\hat{A}^{\dagger}_{k_1,i_1}(0)\right]\right\rangle_{0}=\notag\\
		=&\frac{1}{(2\pi)^2}\bigg\{-\delta_{i_1,i_2}\sum_{\zeta_1,\zeta_2}\Gamma_{\zeta_1,\zeta_2,1}\sum_{\zeta_3,\zeta_4}\Gamma^*_{\zeta_3,\zeta_4,-1}\bigg[	\int_{-\infty}^{+\infty} dk_1\,e^{-i E_{A_{i_1}}(k_1)t'/\hbar}\frac{e^{\beta E_{A_{i_1}}(k_1)}}{1+e^{\beta E_{A_{i_1}}(k_1)}}a_{\zeta_1,i_1}a_{\zeta_4,i_1}e^{ik_1\left[L-(\delta_{\zeta_1,\zeta_2}+\delta_{\zeta_3,\zeta_4})\frac{\xi}{2}\right]}\cdot\notag\\
		&\cdot\int_{-\infty}^{+\infty} dk_2\,e^{-i E_{A_{i_1}}(k_2)t'/\hbar}\frac{e^{\beta E_{A_{i_1}}(k_2)}}{1+e^{\beta E_{A_{i_1}}(k_2)}}a_{\zeta_2,i_1}a_{\zeta_3,i_1}e^{ik_2\left[L-(\delta_{\zeta_1,\zeta_2}+\delta_{\zeta_3,\zeta_4})\frac{\xi}{2}\right]}\bigg]+\notag\\
		\notag\\
		&+\int_{-\infty}^{+\infty} dk_1\,e^{-i E_{A_{i_1}}(k_1)t'/\hbar}\frac{e^{\beta E_{A_{i_1}}(k_1)}}{1+e^{\beta E_{A_{i_1}}(k_1)}}\int_{-\infty}^{+\infty} dk_2\,e^{-i E_{A_{i_2}}(k_2)t'/\hbar}\frac{e^{\beta E_{A_{i_2}}(k_2)}}{1+e^{\beta E_{A_{i_2}}(k_2)}}\cdot\notag\\
		&\cdot\bigg[\sum_{\zeta_1,\zeta_2}\bigg(\Gamma_{\zeta_1,\zeta_2,1}a_{\zeta_1,i_1}a_{\zeta_2,i_2}e^{ik_1\left[\frac{L}{2}-\delta_{\zeta_1,\zeta_2}\frac{\xi}{2}\right]}e^{ik_2\left[\frac{L}{2}+\delta_{\zeta_1,\zeta_2}\frac{\xi}{2}\right]}-\Gamma_{\zeta_1,\zeta_2,1}a_{\zeta_1,i_2}a_{\zeta_2,i_1}e^{ik_2\left[\frac{L}{2}-\delta_{\zeta_1,\zeta_2}\frac{\xi}{2}\right]}e^{ik_1\left[\frac{L}{2}+\delta_{\zeta_1,\zeta_2}\frac{\xi}{2}\right]}\bigg)\cdot\notag\\
		&\cdot\sum_{\zeta_3,\zeta_4}\bigg(\Gamma^*_{\zeta_3,\zeta_4,-1}a_{\zeta_3,i_1}a_{\zeta_4,i_2}e^{-ik_1\left[-\frac{L}{2}-\delta_{\zeta_3,\zeta_4}\frac{\xi}{2}\right]}e^{-ik_2\left[-\frac{L}{2}+\delta_{\zeta_3,\zeta_4}\frac{\xi}{2}\right]}-\Gamma^*_{\zeta_3,\zeta_4,-1}a_{\zeta_3,i_2}a_{\zeta_4,i_1}e^{-ik_2\left[-\frac{L}{2}-\delta_{\zeta_3,\zeta_4}\frac{\xi}{2}\right]}e^{-ik_1\left[-\frac{L}{2}+\delta_{\zeta_3,\zeta_4}\frac{\xi}{2}\right]}\bigg)\bigg]\bigg\}.
	\end{align}
\normalsize

In the previous expression, the building block is represented by the integral
\footnotesize
\begin{align}
   \int_{-\infty}^{+\infty} dk\,e^{-i E_{A_{i_1/i_2}}(k)t'/\hbar}\frac{e^{\beta E_{A_{i_1/i_2}}(k)}}{1+e^{\beta E_{A_{i_1/i_2}}(k)}}e^{ik\left[L+(\pm\delta_{\zeta_1,\zeta_2}\pm\delta_{\zeta_3,\zeta_4})\frac{\xi}{2}\right]},
\end{align}
\normalsize
where all the combinations of signs $\pm$ are possibile. Let the generic energy dispersion be written as $E_{A_{i_1/i_2}}(k)= f_{i_1/i_2}+\rho_{i_1/i_2}\hbar v_F k-\mu$, with $\rho_{i_1/i_2}=\pm1$ and $f_{i_1/i_2}=\pm f$, then it can be computed
\footnotesize
\begin{align}
   &\int_{-\infty}^{+\infty} dk\,e^{-i E_{A_{i_1/i_2}}(k)t'/\hbar}\frac{e^{\beta E_{A_{i_1/i_2}}(k)}}{1+e^{\beta E_{A_{i_1/i_2}}(k)}}e^{ik\left[L+(\pm\delta_{\zeta_1,\zeta_2}\pm\delta_{\zeta_3,\zeta_4})\frac{\xi}{2}\right]}=\notag\\
   &=\exp\left\{{\left[\left(L+(\pm\delta_{\zeta_1,\zeta_2}\pm\delta_{\zeta_3,\zeta_4})\frac{\xi}{2}\right)i-\rho_{i_1/i_2}\xi\right]\frac{(\mu-f_{i_1/i_2})}{\rho_{i_1/i_2}\hbar v_F}}\right\}\frac{1}{\sinh{[\frac{\pi}{\hbar\beta v_F}}(L\pm(\delta_{\zeta_1,\zeta_2}\pm\delta_{\zeta_3,\zeta_4})\frac{\xi}{2}-\rho_{i_1/i_2}v_F t'+i\rho_{i_1/i_2}\xi)]}\cdot\notag\\
   &\cdot\left(\frac{\pi i}{\rho_{i_1/i_2}\beta\hbar v_F}\right).
\end{align}

\normalsize    
Inserting these results in Eq. (\ref{eqn:current2}) and distinguishing explicitly the cases $i_1=i_2$ and $i_1\neq i_2$ in the sum over $i_1i_2$, the current $I^r(t)$ reads as follows
	\footnotesize
	\begin{align}
		I^r(t)=&\frac{8e}{(2\pi\hbar)^2}\text{Im}\bigg\{\int_{-\infty}^{+\infty}dt'\Theta(t')\sum_{i_1,i_2}\sum_{\zeta_1,\zeta_2}\sum_{\zeta_3,\zeta_4}\Gamma_{\zeta_1,\zeta_2,+}\Gamma^*_{\zeta_3,\zeta_4,-}\notag\\
		&\bigg[-\delta_{i_1,i_2}\text{Im}\bigg\{a_{\zeta_1, i_1}a_{\zeta_4, i_1}a_{\zeta_2, i_1}a_{\zeta_3, i_1}\exp{\left\{\left[\left(L-(\delta_{\zeta_1,\zeta_2}+\delta_{\zeta_3,\zeta_4})\frac{\xi}{2}\right)i-\rho_{i_1}\lambda\right]\frac{(\mu-f_{i_1})}{\rho_{i_1}\hbar v_F}\right\}}\cdot\notag\\
		&\cdot\frac{1}{\sinh{[\frac{\pi}{\hbar\beta v_F}}(L-(\delta_{\zeta_1,\zeta_2}+\delta_{\zeta_3,\zeta_4})\frac{\xi}{2}-\rho_{i_1}v_F t'+i\rho_{i_1}\lambda)]}\exp{\left\{\left[\left(L+(\delta_{\zeta_1,\zeta_2}+\delta_{\zeta_3,\zeta_4})\frac{\xi}{2}\right)i-\rho_{i_1}\lambda\right]\frac{(\mu-f_{i_1})}{\rho_{i_1}\hbar v_F}\right\}}\cdot\notag\\ 
		&\cdot\frac{1}{\sinh{[\frac{\pi}{\hbar\beta v_F}}(L+(\delta_{\zeta_1,\zeta_2}+\delta_{\zeta_3,\zeta_4})\frac{\xi}{2}-\rho_{i_1}v_F t'+i\rho_{i_1}\lambda)]}\left(\frac{\pi i}{\rho_{i_1}\beta\hbar v_F}\right)^2\bigg\}+\notag\\
		&\notag\\
		&+\delta_{i_1,i_2}\text{Im}\bigg\{a_{\zeta_1, i_1}a_{\zeta_2, i_2}a_{\zeta_3, i_1}a_{\zeta_4, i_2}\exp{\left\{\left[\left(L+(-\delta_{\zeta_1,\zeta_2}+\delta_{\zeta_3,\zeta_4})\frac{\xi}{2}\right)i-\rho_{i_1}\lambda\right]\frac{(\mu-f_{i_1})}{\rho_{i_1}\hbar v_F}\right\}}\cdot\notag\\
		&\cdot\frac{1}{\sinh{[\frac{\pi}{\hbar\beta v_F}}(L+(-\delta_{\zeta_1,\zeta_2}+\delta_{\zeta_3,\zeta_4})\frac{\xi}{2}-\rho_{i_1}v_F t'+i\rho_{i_1}\lambda)]}\exp{\left\{\left[\left(L+(\delta_{\zeta_1,\zeta_2}-\delta_{\zeta_3,\zeta_4})\frac{\xi}{2}\right)i-\rho_{i_2}\lambda\right]\frac{(\mu-f_{i_2})}{\rho_{i_2}\hbar v_F}\right\}}\cdot\notag\\ 
		&\cdot\frac{1}{\sinh{[\frac{\pi}{\hbar\beta v_F}}(L+(\delta_{\zeta_1,\zeta_2}-\delta_{\zeta_3,\zeta_4})\frac{\xi}{2}-\rho_{i_2}v_F t'+i\rho_{i_2}\lambda)]}\left(\frac{\pi i}{\rho_{i_1}\beta\hbar v_F}\right)\left(\frac{\pi i}{\rho_{i_2}\beta\hbar v_F}\right)\bigg\}+\notag\\
		&\notag\\
		&+\delta_{i_1,-i_2}\text{Im}\bigg\{a_{\zeta_1, i_1}a_{\zeta_2, i_2}a_{\zeta_3, i_1}a_{\zeta_4, i_2}\exp{\left\{\left[\left(L+(-\delta_{\zeta_1,\zeta_2}+\delta_{\zeta_3,\zeta_4})\frac{\xi}{2}\right)i-\rho_{i_1}\lambda\right]\frac{(\mu-f_{i_1})}{\rho_{i_1}\hbar v_F}\right\}}\cdot\notag\\
		&\cdot\frac{1}{\sinh{[\frac{\pi}{\hbar\beta v_F}}(L+(-\delta_{\zeta_1,\zeta_2}+\delta_{\zeta_3,\zeta_4})\frac{\xi}{2}-\rho_{i_1}v_F t'+i\rho_{i_1}\lambda)]}\exp{\left\{\left[\left(L+(\delta_{\zeta_1,\zeta_2}-\delta_{\zeta_3,\zeta_4})\frac{\xi}{2}\right)i-\rho_{i_2}\lambda\right]\frac{(\mu-f_{i_2})}{\rho_{i_2}\hbar v_F}\right\}}\cdot\notag\\ 
		&\cdot\frac{1}{\sinh{[\frac{\pi}{\hbar\beta v_F}}(L+(\delta_{\zeta_1,\zeta_2}-\delta_{\zeta_3,\zeta_4})\frac{\xi}{2}-\rho_{i_2}v_F t'+i\rho_{i_2}\lambda)]}\left(\frac{\pi i}{\rho_{i_1}\beta\hbar v_F}\right)\left(\frac{\pi i}{\rho_{i_2}\beta\hbar v_F}\right)\bigg\}+\notag\\
		&\notag\\
		&-\delta_{i_1,-i_2}\text{Im}\bigg\{a_{\zeta_1, i_1}a_{\zeta_2, i_2}a_{\zeta_3, i_2}a_{\zeta_4, i_1}\exp{\left\{\left[\left(L+(-\delta_{\zeta_1,\zeta_2}-\delta_{\zeta_3,\zeta_4})\frac{\xi}{2}\right)i-\rho_{i_1}\lambda\right]\frac{(\mu-f_{i_1})}{\rho_{i_1}\hbar v_F}\right\}}\cdot\notag\\
		&\cdot\frac{1}{\sinh{[\frac{\pi}{\hbar\beta v_F}}(L+(-\delta_{\zeta_1,\zeta_2}-\delta_{\zeta_3,\zeta_4})\frac{\xi}{2}-\rho_{i_1}v_F t'+i\rho_{i_1}\lambda)]}\exp{\left\{\left[\left(L+(\delta_{\zeta_1,\zeta_2}+\delta_{\zeta_3,\zeta_4})\frac{\xi}{2}\right)i-\rho_{i_2}\lambda\right]\frac{(\mu-f_{i_2})}{\rho_{i_2}\hbar v_F}\right\}}\cdot\notag\\ 
		&\cdot\frac{1}{\sinh{[\frac{\pi}{\hbar\beta v_F}}(L+(\delta_{\zeta_1,\zeta_2}+\delta_{\zeta_3,\zeta_4})\frac{\xi}{2}-\rho_{i_2}v_F t'+i\rho_{i_2}\lambda)]}\left(\frac{\pi i}{\rho_{i_1}\beta\hbar v_F}\right)\left(\frac{\pi i}{\rho_{i_2}\beta\hbar v_F}\right)\bigg\}+\notag\\
		&\notag\\
		&-\delta_{i_1,-i_2}\text{Im}\bigg\{a_{\zeta_1, i_2}a_{\zeta_2, i_1}a_{\zeta_3, i_1}a_{\zeta_4, i_2}\exp{\left\{\left[\left(L+(\delta_{\zeta_1,\zeta_2}+\delta_{\zeta_3,\zeta_4})\frac{\xi}{2}\right)i-\rho_{i_1}\lambda\right]\frac{(\mu-f_{i_1})}{\rho_{i_1}\hbar v_F}\right\}}\cdot\notag\\
		&\cdot\frac{1}{\sinh{[\frac{\pi}{\hbar\beta v_F}}(L+(\delta_{\zeta_1,\zeta_2}+\delta_{\zeta_3,\zeta_4})\frac{\xi}{2}-\rho_{i_1}v_F t'+i\rho_{i_1}\lambda)]}\exp{\left\{\left[\left(L+(-\delta_{\zeta_1,\zeta_2}-\delta_{\zeta_3,\zeta_4})\frac{\xi}{2}\right)i-\rho_{i_2}\lambda\right]\frac{(\mu-f_{i_2})}{\rho_{i_2}\hbar v_F}\right\}}\cdot\notag\\ 
		&\cdot\frac{1}{\sinh{[\frac{\pi}{\hbar\beta v_F}}(L+(-\delta_{\zeta_1,\zeta_2}-\delta_{\zeta_3,\zeta_4})\frac{\xi}{2}-\rho_{i_2}v_F t'+i\rho_{i_2}\lambda)]}\left(\frac{\pi i}{\rho_{i_1}\beta\hbar v_F}\right)\left(\frac{\pi i}{\rho_{i_2}\beta\hbar v_F}\right)\bigg\}+\notag\\
		&\notag\\
		&+\delta_{i_1,-i_2}\text{Im}\bigg\{a_{\zeta_1, i_2}a_{\zeta_2, i_1}a_{\zeta_3, i_2}a_{\zeta_4, i_1}\exp{\left\{\left[\left(L+(\delta_{\zeta_1,\zeta_2}-\delta_{\zeta_3,\zeta_4})\frac{\xi}{2}\right)i-\rho_{i_1}\lambda\right]\frac{(\mu-f_{i_1})}{\rho_{i_1}\hbar v_F}\right\}}\cdot\notag\\
		&\cdot\frac{1}{\sinh{[\frac{\pi}{\hbar\beta v_F}}(L+(\delta_{\zeta_1,\zeta_2}-\delta_{\zeta_3,\zeta_4})\frac{\xi}{2}-\rho_{i_1}v_F t'+i\rho_{i_1}\lambda)]}\exp{\left\{\left[\left(L+(-\delta_{\zeta_1,\zeta_2}+\delta_{\zeta_3,\zeta_4})\frac{\xi}{2}\right)i-\rho_{i_2}\lambda\right]\frac{(\mu-f_{i_2})}{\rho_{i_2}\hbar v_F}\right\}}\cdot\notag\\ 
		&\cdot\frac{1}{\sinh{[\frac{\pi}{\hbar\beta v_F}}(L+(-\delta_{\zeta_1,\zeta_2}+\delta_{\zeta_3,\zeta_4})\frac{\xi}{2}-\rho_{i_2}v_F t'+i\rho_{i_2}\lambda)]}\left(\frac{\pi i}{\rho_{i_1}\beta\hbar v_F}\right)\left(\frac{\pi i}{\rho_{i_2}\beta\hbar v_F}\right)\bigg\}\bigg]\bigg\}.\label{eqn:currentApp}
	\end{align}
	\normalsize
	
The summation over $i_1,i_2$ runs over 10 terms that can be divided into three cases:
\begin{enumerate}
    \item $i_1=i_2$, including $i_1,i_2=1,1\quad2,2\quad3,3\quad4,4$;
    \item $i_1\neq i_2$, with $\hat{A}_{k_1,i_1}$ and $\hat{A}_{k_2,i_2}$ being both right-movers or both left-movers, including $i_1,i_2=1,2\quad3,4$;
    \item $i_1\neq i_2$, with $\hat{A}_{k_1,i_1}$ and $\hat{A}_{k_2,i_2}$ having opposite directions of propagation, including $i_1,i_2=1,3\quad1,4\quad2,3\quad2,4$.
\end{enumerate}
Table \ref{table1} makes explicit the three cases, by looking at Eq. (\ref{eqn:En.A2}).
\footnotesize
\begin{table}[hbt]
\centering
\begin{tabular}{lccccr}
\toprule
 & & $f_{i_1}$ & $f_{i_2}$ & $\rho_{i_1}$ & $\rho_{i_2}$ \\
\midrule
(i)   & 1,\,1 & $-f$ & $-f$ & -1 & -1\\
      & 2,\,2 & $f$ & $f$ & -1 & -1\\
      & 3,\,3 & $-f$ & $-f$ & 1 & 1\\
      & 4,\,4 & $f$ & $f$ & 1 & 1\\
(ii)  & 1,\,2 & $-f$ & $f$ & -1 & -1\\
      & 3,\,4 & $-f$ & $f$ & 1 & 1\\
(iii) & 1,\,3 & $-f$ & $-f$ & -1 & 1\\
      & 1,\,4 & $-f$ & $f$ & -1 & 1\\
      & 2,\,3 & $f$ & $-f$ & -1 & 1\\
      & 2,\,4 & $f$ & $f$ & -1 & 1\\
\bottomrule
\end{tabular}
\caption{Details concerning the ten terms of the summation over $i_1,i_2$ in Eq. (\ref{eqn:currentApp}).}
\label{table1}
\end{table}
\normalsize

The next step consists in developing the summations over $\zeta_1,\zeta_2$ and $\zeta_3,\zeta_4$, remembering that the possible values are only those listed in (\ref{eqn:zeta}). Since some of the $a_{\zeta,i}$ are zero, the corresponding addends of the sum will not contribute to the current. The expression we obtain, using the dimensionless quantities introduced in the main text, is
\footnotesize
\begin{align*}
    &I^r(t)=\frac{-2e\Delta\Tilde{T}^2\Gamma^2}{\pi^2\hbar^3v_F^2}\text{Im}\Bigg\{e^{-i(\omega_J t+\varphi^0_r-\varphi^0_l)}\int_0^{+\infty}ds e^{is\Tilde{V}}\Bigg\{\\
    &\Tilde{f}_T^2e^{-i2k_F L}\cos{\left(\pi\frac{\phi}{\phi_0}\right)}\text{Im}\Bigg[\frac{1}{2}\left(e^{\left(2\Tilde{L}i+2\right)\left(-\Tilde{\mu}-\Tilde{f}\right)}+e^{\left(2\Tilde{L}i+2\right)\left(-\Tilde{\mu}+\Tilde{f}\right)}\right)\\
    &\left(\frac{1}{\sinh^2{\left[\Tilde{T}\left(\Tilde{L}+s-i\right)\right]}}-\frac{1}{\sinh{\left[\Tilde{T}\left(\Tilde{L}-1+s-i\right)\right]}\sinh{\left[\Tilde{T}\left(\Tilde{L}+1+s-i\right)\right]}}\right)+e^{-\left(2\Tilde{L}i+2\right)\Tilde{\mu}}\frac{1}{\sinh^2{\left[\Tilde{T}\left(\Tilde{L}+s-i\right)\right]}}+\\
    &-\frac{1}{2}\left(e^{2i\Tilde{f}}e^{-\left(2\Tilde{L}i+2\right)\Tilde{\mu}}+e^{-2i\Tilde{f}}e^{-\left(2\Tilde{L}i+2\right)\Tilde{\mu}}\right)\frac{1}{\sinh{\left[\Tilde{T}\left(\Tilde{L}+1+s-i\right)\right]}\sinh{\left[\Tilde{T}\left(\Tilde{L}-1+s-i\right)\right]}}\Bigg]+\\
    &+\Tilde{f}_T^2e^{i2k_F L}\cos{\left(\pi\frac{\phi}{\phi_0}\right)}\text{Im}\Bigg[\frac{1}{2}\left(e^{\left(2\Tilde{L}i-2\right)\left(\Tilde{\mu}+\Tilde{f}\right)}+e^{\left(2\Tilde{L}i-2\right)\left(\Tilde{\mu}-\Tilde{f}\right)}\right)\\
    &\left(\frac{1}{\sinh^2{\left[\Tilde{T}\left(\Tilde{L}-s+i\right)\right]}}-\frac{1}{\sinh{\left[\Tilde{T}\left(\Tilde{L}-1-s+i\right)\right]}\sinh{\left[\Tilde{T}\left(\Tilde{L}+1-s+i\right)\right]}}\right)+\\
    &+e^{\left(2\Tilde{L}i-2\right)\Tilde{\mu}}\frac{1}{\sinh^2{\left[\Tilde{T}\left(\Tilde{L}-s+i\right)\right]}}-\frac{1}{2}e^{-2i\Tilde{f}}e^{\left(2\Tilde{L}i-2\right)\Tilde{\mu}}\frac{1}{\sinh{\left[\Tilde{T}\left(\Tilde{L}-1-s+i\right)\right]}\sinh{\left[\Tilde{T}\left(\Tilde{L}+1-s+i\right)\right]}}+\\
    &-\frac{1}{2}e^{2i\Tilde{f}}e^{\left(2\Tilde{L}i-2\right)\Tilde{\mu}}\frac{1}{\sinh{\left[\Tilde{T}\left(\Tilde{L}-1-s+i\right)\right]}\sinh{\left[\Tilde{T}\left(\Tilde{L}+1-s+i\right)\right]}}\Bigg]+\\
    &-\frac{1}{2}\cos{\left(\pi\frac{\phi}{\phi_0}\right)}\text{Im}\Bigg[\bigg(e^{\left(\Tilde{L}i+1\right)\left(-\Tilde{\mu}-\Tilde{f}\right)}e^{\left(\Tilde{L}i-1\right)\left(\Tilde{\mu}+\Tilde{f}\right)}+e^{\left(\Tilde{L}i+1\right)\left(-\Tilde{\mu}+\Tilde{f}\right)}e^{\left(\Tilde{L}i-1\right)\left(\Tilde{\mu}-\Tilde{f}\right)}+e^{\left(\Tilde{L}i+1\right)\left(-\Tilde{\mu}-\Tilde{f}\right)}e^{\left(\Tilde{L}i-1\right)\left(\Tilde{\mu}-\Tilde{f}\right)}+\\
    &+e^{\left(\Tilde{L}i+1\right)\left(-\Tilde{\mu}+\Tilde{f}\right)}e^{\left(\Tilde{L}i-1\right)\left(\Tilde{\mu}+\Tilde{f}\right)}\bigg)\frac{1}{\sinh{\left[\Tilde{T}\left(\Tilde{L}+s-i\right)\right]}\sinh{\left[\Tilde{T}\left(\Tilde{L}-s+i\right)\right]}}\Bigg]+\\
    &-2\Tilde{f}_T f_C e^{i2k_F L}\sin{\left(\frac{\pi}{2}\frac{\phi}{\phi_0}\right)}\text{Im}\Bigg[\left(e^{i\Tilde{f}}-e^{-i\Tilde{f}}\right)e^{\left(2\Tilde{L}i-2\right)\Tilde{\mu}}\frac{1}{\sinh{\left[\Tilde{T}   \left(\Tilde{L}+\frac{1}{2}-s+i\right)\right]}\sinh{\left[\Tilde{T}\left(\Tilde{L}-\frac{1}{2}-s+i\right)\right]}}\Bigg]+\\
    &+2\Tilde{f}_T f_C e^{-i2k_F L}\sin{\left(\frac{\pi}{2}\frac{\phi}{\phi_0}\right)}\text{Im}\Bigg[\left(e^{i\Tilde{f}}-e^{-i\Tilde{f}}\right)e^{-\left(2\Tilde{L}i+2\right)\Tilde{\mu}}\frac{1}{\sinh{\left[\Tilde{T}\left(\Tilde{L}-\frac{1}{2}+s-i\right)\right]}\sinh{\left[\Tilde{T}\left(\Tilde{L}+\frac{1}{2}+s-i\right)\right]}}\Bigg]+\\
    &+\Tilde{f}_T^2e^{i(-k_F L2)}\text{Im}\Bigg[\frac{1}{2}\left(e^{\left(2\Tilde{L}i+2\right)\left(-\Tilde{\mu}-\Tilde{f}\right)}+e^{\left(2\Tilde{L}i+2\right)\left(-\Tilde{\mu}+\Tilde{f}\right)}\right)\\
    &\left(\frac{1}{\sinh^2{\left[\Tilde{T}\left(\Tilde{L}+s-i\right)\right]}}-\frac{1}{\sinh{\left[\Tilde{T}\left(\Tilde{L}-1+s-i\right)\right]}\sinh{\left[\Tilde{T}\left(\Tilde{L}+1+s-i\right)\right]}}\right)-e^{-\left(2\Tilde{L}i+2\right)\Tilde{\mu}}\frac{1}{\sinh^2{\left[\Tilde{T}\left(\Tilde{L}+s-i\right)\right]}}+\\
    &+\frac{1}{2}\left(e^{2i\Tilde{f}}e^{-\left(2\Tilde{L}i+2\right)\Tilde{\mu}}+e^{-2i\Tilde{f}}e^{-\left(2\Tilde{L}i+2\right)\Tilde{\mu}}\right)\frac{1}{\sinh{\left[\Tilde{T}\left(\Tilde{L}+1+s-i\right)\right]}\sinh{\left[\Tilde{T}\left(\Tilde{L}-1+s-i\right)\right]}}\Bigg]+\\
    &+( f_C^2e^{i(-k_F L2)})\text{Im}\Bigg[e^{-\left(2\Tilde{L}i+2\right)\Tilde{\mu}}\frac{1}{\sinh^2{\left[\Tilde{T}\left(\Tilde{L}+s-i\right)\right]}}\Bigg]+
\end{align*}
\begin{align*}
    &+(\Tilde{f}_T^2e^{i(k_F L2)})\text{Im}\Bigg[\frac{1}{2}\left(e^{\left(2\Tilde{L}i-2\right)\left(\Tilde{\mu}+\Tilde{f}\right)}+e^{\left(2\Tilde{L}i-2\right)\left(\Tilde{\mu}-\Tilde{f}\right)}\right)\\
    &\left(\frac{1}{\sinh^2{\left[\Tilde{T}\left(\Tilde{L}-s+i\right)\right]}}-\frac{1}{\sinh{\left[\Tilde{T}\left(\Tilde{L}-1-s+i\right)\right]}\sinh{\left[\Tilde{T}\left(\Tilde{L}+1-s+i\right)\right]}}\right)+\\
    &-e^{\left(2\Tilde{L}i-2\right)\Tilde{\mu}}\frac{1}{\sinh^2{\left[\Tilde{T}\left(\Tilde{L}-s+i\right)\right]}}+\frac{1}{2}e^{-2i\Tilde{f}}e^{\left(2\Tilde{L}i-2\right)\Tilde{\mu}}\frac{1}{\sinh{\left[\Tilde{T}\left(\Tilde{L}-1-s+i\right)\right]}\sinh{\left[\Tilde{T}\left(\Tilde{L}+1-s+i\right)\right]}}+\\
    &+\frac{1}{2}e^{2i\Tilde{f}}e^{\left(2\Tilde{L}i-2\right)\Tilde{\mu}}\frac{1}{\sinh{\left[\Tilde{T}\left(\Tilde{L}-1-s+i\right)\right]}\sinh{\left[\Tilde{T}\left(\Tilde{L}+1-s+i\right)\right]}}\Bigg]+\\
    &-(\Tilde{f}_T^2 f_C^2)\text{Im}\Bigg[\left(e^{\left(\Tilde{L}i+1\right)\left(-\Tilde{\mu}-\Tilde{f}\right)}e^{\left(\Tilde{L}i-1\right)\left(\Tilde{\mu}+\Tilde{f}\right)}+e^{\left(\Tilde{L}i+1\right)\left(-\Tilde{\mu}+\Tilde{f}\right)}e^{\left(\Tilde{L}i-1\right)\left(\Tilde{\mu}-\Tilde{f}\right)}\right)\frac{1}{\sinh{\left[\Tilde{T}\left(\Tilde{L}+s-i\right)\right]}\sinh{\left[\Tilde{T}\left(\Tilde{L}-s+i\right)\right]}}\Bigg]+\\
    &-\frac{1}{2}\text{Im}\Bigg[\bigg(e^{\left(\Tilde{L}i+1\right)\left(-\Tilde{\mu}-\Tilde{f}\right)}e^{\left(\Tilde{L}i-1\right)\left(\Tilde{\mu}+\Tilde{f}\right)}+e^{\left(\Tilde{L}i+1\right)\left(-\Tilde{\mu}+\Tilde{f}\right)}e^{\left(\Tilde{L}i-1\right)\left(\Tilde{\mu}-\Tilde{f}\right)}-e^{\left(\Tilde{L}i+1\right)\left(-\Tilde{\mu}-\Tilde{f}\right)}e^{\left(\Tilde{L}i-1\right)\left(\Tilde{\mu}-\Tilde{f}\right)}+\\
    &+e^{\left(\Tilde{L}i+1\right)\left(-\Tilde{\mu}+\Tilde{f}\right)}e^{\left(\Tilde{L}i-1\right)\left(\Tilde{\mu}+\Tilde{f}\right)}\bigg)\frac{1}{\sinh{\left[\Tilde{T}\left(\Tilde{L}+s-i\right)\right]}\sinh{\left[\Tilde{T}\left(\Tilde{L}-s+i\right)\right]}}\Bigg]+\\
    &+f_C^2e^{i2k_F L}\text{Im}\Bigg[e^{\left(2\Tilde{L}i-2\right)\Tilde{\mu}}\frac{1}{\sinh^2{\left[\Tilde{T}\left(\Tilde{L}-s+i\right)\right]}}\Bigg]\Bigg\}\Bigg\}.
\end{align*}
\normalsize

At this stage, it is already possible to visualise the structure of Eq. (\ref{eqn:currentflux}).

The last step is the evaluation of the eighteen integrals in the previous expression. In particular
\footnotesize
\begin{align}
    I_1&=\int_0^{+\infty}ds\,e^{i\tilde{V}s}\text{Im}\frac{\tilde{T}^2}{\sinh{\left[\tilde{T}\left(\tilde{L}-s+i\right)\right]}\sinh{\left[\tilde{T}\left(\tilde{L}+s-i\right)\right]}},\label{eqn:primoint}\\
    I_2&=\int_0^{+\infty}ds\,e^{i\tilde{V}s}\text{Re}\frac{\tilde{T}^2}{\sinh{\left[\tilde{T}\left(\tilde{L}-s+i\right)\right]}\sinh{\left[\tilde{T}\left(\tilde{L}+s-i\right)\right]}},\\
    I_3&=\int_0^{+\infty}ds\,e^{i\tilde{V}s}\text{Im}\frac{\tilde{T}^2}{\sinh^2{\left[\tilde{T}\left(\tilde{L}-s+i\right)\right]}},\\
    I_4&=\int_0^{+\infty}ds\,e^{i\tilde{V}s}\text{Re}\frac{\tilde{T}^2}{\sinh^2{\left[\tilde{T}\left(\tilde{L}-s+i\right)\right]}},\\
    I_5&=\int_0^{+\infty}ds\,e^{i\tilde{V}s}\text{Im}\frac{\tilde{T}^2}{\sinh^2{\left[\tilde{T}\left(\tilde{L}+s-i\right)\right]}},\\
    I_6&=\int_0^{+\infty}ds\,e^{i\tilde{V}s}\text{Re}\frac{\tilde{T}^2}{\sinh^2{\left[\tilde{T}\left(\tilde{L}+s-i\right)\right]}},\\
    I_7&=\int_0^{+\infty}ds\,e^{i\tilde{V}s}\text{Im}\frac{\tilde{T}^2}{\sinh^4{\left[\tilde{T}\left(\tilde{L}-s+i\right)\right]}},\\
    I_8&=\int_0^{+\infty}ds\,e^{i\tilde{V}s}\text{Re}\frac{\tilde{T}^2}{\sinh^4{\left[\tilde{T}\left(\tilde{L}-s+i\right)\right]}},\\
    I_9&=\int_0^{+\infty}ds\,e^{i\tilde{V}s}\text{Im}\frac{\tilde{T}^2}{\sinh^4{\left[\tilde{T}\left(\tilde{L}+s-i\right)\right]}},\\
    I_{10}&=\int_0^{+\infty}ds\,e^{i\tilde{V}s}\text{Re}\frac{\tilde{T}^2}{\sinh^4{\left[\tilde{T}\left(\tilde{L}+s-i\right)\right]}},\label{eqn:ultimoint}\\
    I_{11}&=\int_0^{+\infty}ds\,e^{i\tilde{V}s}\text{Im}\frac{\tilde{T}^2}{\sinh{\left[\tilde{T}\left(\tilde{L}-1-s+i\right)\right]}\sinh{\left[\tilde{T}\left(\tilde{L}+1-s+i\right)\right]}},\\
    I_{12}&=\int_0^{+\infty}ds\,e^{i\tilde{V}s}\text{Re}\frac{\tilde{T}^2}{\sinh{\left[\tilde{T}\left(\tilde{L}-1-s+i\right)\right]}\sinh{\left[\tilde{T}\left(\tilde{L}+1-s+i\right)\right]}},\\
    I_{13}&=\int_0^{+\infty}ds\,e^{i\tilde{V}s}\text{Im}\frac{\tilde{T}^2}{\sinh{\left[\tilde{T}\left(\tilde{L}-\frac{1}{2}-s+i\right)\right]}\sinh{\left[\tilde{T}\left(\tilde{L}+\frac{1}{2}-s+i\right)\right]}},\\
    I_{14}&=\int_0^{+\infty}ds\,e^{i\tilde{V}s}\text{Re}\frac{\tilde{T}^2}{\sinh{\left[\tilde{T}\left(\tilde{L}-\frac{1}{2}-s+i\right)\right]}\sinh{\left[\tilde{T}\left(\tilde{L}+\frac{1}{2}-s+i\right)\right]}},\\
    I_{15}&=\int_0^{+\infty}ds\,e^{i\tilde{V}s}\text{Im}\frac{\tilde{T}^2}{\sinh{\left[\tilde{T}\left(\tilde{L}-1+s-i\right)\right]}\sinh{\left[\tilde{T}\left(\tilde{L}+1+s-i\right)\right]}},\\
    I_{16}&=\int_0^{+\infty}ds\,e^{i\tilde{V}s}\text{Re}\frac{\tilde{T}^2}{\sinh{\left[\tilde{T}\left(\tilde{L}-1+s-i\right)\right]}\sinh{\left[\tilde{T}\left(\tilde{L}+1+s-i\right)\right]}},\\
    I_{17}&=\int_0^{+\infty}ds\,e^{i\tilde{V}s}\text{Im}\frac{\tilde{T}^2}{\sinh{\left[\tilde{T}\left(\tilde{L}-\frac{1}{2}+s-i\right)\right]}\sinh{\left[\tilde{T}\left(\tilde{L}+\frac{1}{2}+s-i\right)\right]}},\\
    I_{18}&=\int_0^{+\infty}ds\,e^{i\tilde{V}s}\text{Re}\frac{\tilde{T}^2}{\sinh{\left[\tilde{T}\left(\tilde{L}-\frac{1}{2}+s-i\right)\right]}\sinh{\left[\tilde{T}\left(\tilde{L}+\frac{1}{2}+s-i\right)\right]}}.
\end{align}
\normalsize

Only those in Eqs. (\ref{eqn:primoint})-(\ref{eqn:ultimoint}) are independent. Indeed, up to the second order in $\xi/L$, we have
\footnotesize
\begin{align*}
   I_{11}&\approx I_7+I_3,\\
   I_{12}&\approx I_8+I_4,\\
   I_{13}&\approx \frac{1}{4}I_7+I_3,\\
   I_{14}&\approx \frac{1}{4}I_8+I_4,\\
   I_{15}&\approx I_9+I_5,\\
   I_{16}&\approx I_{10}+I_6,\\
   I_{17}&\approx \frac{1}{4}I_9+I_5,\\
   I_{18}&\approx \frac{1}{4}I_{10}+I_6.
\end{align*}
\normalsize

Some of the independent integrals are already present in \cite{loss}, the other ones have been evaluated under the same assumptions and up to the same order in $\xi$. We obtain
\footnotesize
\begin{align*}
    I_1&\approx-\frac{\pi\tilde{T}e^{\left(i\tilde{V}\tilde{L}-\tilde{V}\right)}}{\sinh{\left(2\tilde{T}\tilde{L}\right)}},\\
    I_2&\approx-i\frac{\pi\tilde{T}e^{\left(i\tilde{V}\tilde{L}-\tilde{V}\right)}}{\sinh{\left(2\tilde{T}\tilde{L}\right)}},\\
    I_3&\approx i\pi\tilde{V}e^{\left(i\tilde{V}\tilde{L}-\tilde{V}\right)},\\
    I_4&\approx -\pi\tilde{V}e^{\left(i\tilde{V}\tilde{L}-\tilde{V}\right)},\\
    I_5&\approx0,\\
    I_6&\approx0,\\
    I_7&\approx-i\frac{\pi}{6}\tilde{V}\left(\tilde{V}^2+4\tilde{T}^2\right)e^{\left(i\tilde{V}\tilde{L}-\tilde{V}\right)},\\
    I_8&\approx\frac{\pi}{6}\tilde{V}\left(\tilde{V}^2+4\tilde{T}^2\right)e^{\left(i\tilde{V}\tilde{L}-\tilde{V}\right)},\\
    I_9&\approx0,\\
    I_{10}&\approx0.
\end{align*}

\normalsize
Once the integration over $s$ is done, we obtain
	\begin{equation}
		I^{r}(t)=\mathcal{C}\,\text{Im}\bigg\{e^{- i(\omega_J t+\varphi^0_r-\varphi^0_l)}\left[\tilde{A}^r_1\cos{\left(\pi\frac{\phi}{\phi_0}\right)}+\tilde{A}^r_2\sin{\left(\frac{\pi}{2}\frac{\phi}{\phi_0}\right)}+\tilde{A}^r_3\right]\bigg\},
	\end{equation}
with $\mathcal{C}=(-2e\Delta\Gamma^2)/(\pi^2\hbar^3v_F^2)$ as defined in the main text and
\small
\begin{subequations}
	\begin{align}
		&\tilde{A}^r_1=\frac{\pi}{6}\,\exp{\left[-2\tilde{\mu}+\tilde{V}(-1+i\tilde{L})\right]}\bigg\{i\,\exp{\left[i\left(2k_FL+2\tilde{L}\tilde{\mu}\right)\right]}\,\tilde{f}_T^2\,\tilde{V}\bigg[6+\left(-6+4\,\tilde{T}^2+\tilde{V}^2\right)\cdot\\
		&\cos{\left(2\tilde{f}\right)}+\left(4\,\tilde{T}^2+\tilde{V}^2\right)\cosh{\left(2\tilde{f}-i2\tilde{f}\tilde{L}\right)}\bigg]+\frac{12\,\tilde{T}}{\sinh{\left(2\tilde{L}\tilde{T}\right)}}\,\cosh{\left(\tilde{f}-i\tilde{f}\tilde{L}\right)}\,\cosh{\left(\tilde{f}+i\tilde{f}\tilde{L}\right)}\bigg\},\notag\\
		&\tilde{A}^r_2=-\frac{\pi f_C\tilde{f}_T\tilde{V}}{6}\exp{\left[i2k_FL+\left(-1+i\tilde{L}\right)\left(\tilde{V}+2\tilde{\mu}\right)\right]}\left(-24+4\tilde{T}^2+\tilde{V}^2\right)\sin{\left(2\tilde{f}\right)},\\
		&\tilde{A}^r_3=\frac{\pi}{6}\,\exp{\left[-2\tilde{\mu}+\tilde{V}(-1+i\tilde{L})\right]}\bigg\{-i\,\exp{\left[2ik_FL+2i\tilde{L}\tilde{\mu}\right]}\,\tilde{V}\bigg[-6\left(f_C^2-\tilde{f}_T^2\right)+\\
		&+\tilde{f}_T^2\left(-6+4\tilde{T}^2+\tilde{V}^2\right)\cos{\left(2\tilde{f}\right)}-\tilde{f}_T^2\left(4\tilde{T}^2+\tilde{V}^2\right)\cosh{\left(2\tilde{f}-2i\tilde{f}\tilde{L}\right)}\bigg]+\notag\\
		&+6\frac{\tilde{T}}{\sinh{\left(2\tilde{L}\tilde{T}\right)}}\bigg[\left(-1+2f_C^2\tilde{f}_T^2\right)\cosh{\left(2\tilde{f}\right)}+\cosh{\left(-2i\tilde{f}\tilde{L}\right)}\bigg]\bigg\}.\notag
	\end{align}\label{eqn:coeffs}
\end{subequations}
\normalsize
\end{document}